\documentclass[british,amssymb,aps,prd,twocolumn,amsmath,floats,floatfix,superscriptaddress,nofootinbib,%
               showkeys,showpacs,groupedaddress]{revtex4-1}
\usepackage{amssymb}
\usepackage{amsmath}
\usepackage{verbatim}
\usepackage{mathrsfs}
\usepackage{amsfonts}
\usepackage{latexsym}
\usepackage{epsfig}
\usepackage{color}
\usepackage[usenames,dvipsnames]{xcolor}
\usepackage{graphicx,subfigure}
\usepackage{units}
\usepackage{envmath}
\usepackage{natbib}
\usepackage{ctable}
\usepackage{soul}
\usepackage[unicode=true,pdfusetitle,
 bookmarks=true,bookmarksnumbered=false,bookmarksopen=false,
 breaklinks=false,pdfborder={0 0 1},backref=false,colorlinks=false]{hyperref}
\usepackage{mathtools}
\begin{document}
\definecolor{orange}{rgb}{0.9,0.45,0}
\def\CovDev{D}
\def\Res{{\mathcal R}}
\def\Gammaflat{\hat \Gamma}
\def\metricflat{\hat \gamma}
\def\Dflat{\hat {\mathcal D}}
\def\part_n{\partial_\perp}
\def\Lie{\mathcal{L}}
\def\A{\mathcal{X}}
\def\Aphi{\A_{\phi}}
\def\hAphi{\hat{\A}_{\phi}}
\def\E{\mathcal{E}}
\def\Ham{\mathcal{H}}
\def\M{\mathcal{M}}
\def\R{\mathcal{R}}
\def\p{\partial}
\def\hg{\hat{\gamma}}
\def\hA{\hat{A}}
\def\hD{\hat{D}}
\def\hE{\hat{E}}
\def\hR{\hat{R}}
\def\hcA{\hat{\mathcal{A}}}
\def\hDelt{\hat{\triangle}}
\def\na{\nabla}
\def\dif{{\rm{d}}}
\def\non{\nonumber}
\newcommand{\erf}{\textrm{erf}}
\renewcommand{\t}{\times}
\long\def\symbolfootnote[#1]#2{\begingroup%
\def\thefootnote{\fnsymbol{footnote}}\footnote[#1]{#2}\endgroup}
\title{Cosmological Inflation in $f(Q, T)$ Gravity}
\author{Maryam Shiravand}
\email{ma\_shiravand@sbu.ac.ir}
\author{Saeed Fakhry}
\email{s\_fakhry@sbu.ac.ir}
\author{Mehrdad Farhoudi}
\email{m-farhoudi@sbu.ac.ir}
 \affiliation{Department of Physics,
              Shahid Beheshti University, Evin, Tehran 19839, Iran}

\date{August 8, 2022}
\begin{abstract}
\noindent
 We study the cosmological inflation within the context
of $f(Q, T)$ gravity, wherein $Q$ is the nonmetricity scalar and
$T$ is the trace of the matter energy-momentum tensor. By choosing
a linear combination of $Q$ and $T$, we first analyze the
realization of an inflationary scenario driven via the geometrical
effects of the linear $f(Q, T)$ gravity and then, we obtain the
modified slow-roll parameters, the scalar and the tensor spectral
indices, and the tensor-to-scalar ratio for the proposed model. In
addition, by choosing three inflationary potentials, i.e. the
power-law, hyperbolic and natural potentials, and by applying the
slow-roll approximations, we calculate these inflationary
observables in the presence of an inflaton scalar field. The
results indicate that by properly restricting the free parameters,
the proposed model provides appropriate predictions that are
consistent with the observational data obtained from the Planck
2018. Also, we specify that the contribution of linear model of
$f(Q,T)$ gravity with the hyperbolic and natural potentials can
impose different restrictions on the parameters of these
potentials. Furthermore, the predictions of natural inflation in
this model are in good agreement with the joint Planck, BK15 and
BAO data, justifying the use of the $f(Q, T)$ gravity.
\end{abstract}

\maketitle

\section{Introduction}
In recent decades, a wide range of theoretical and observational
studies have been conducted to describe the dynamics of the
Universe, and the results appear to be almost in good agreement
with the standard cosmological model, i.e., the $\Lambda$CDM
model, which stems from general relativity
(GR)~\cite{Ferreira:2019xrr}. On the other hand, the observations
related to the cosmic microwave background radiation (CMB) through
various surveys contain very important information about the
formation and evolution of the Universe~\cite{WMAP:2003elm,
WMAP:2006bqn, WMAP:2008lyn, Hinshaw2013,Planck:2018jri,
Planck:2018vyg}. However, some challenging concepts, such as
flatness and horizon problems, remain undesirable within the
framework of the standard cosmological model~\cite{Coley:2019yov}.
To solve such problems, it has been proposed to consider an epoch
of accelerated expansion in the earliest stage of the evolution of
the Universe, known as cosmic inflation~\cite{Starobinsky1980,
guth1981, lidd1982, Albrecht1982}.

The most appropriate way to explain the inflationary framework is
to consider a scalar field known as inflaton, which is managed by
a specific potential. In particular, the quantum fluctuations of
the inflaton can give rise to an inflationary era and explain the
origin of the large-scale structures. In other words, the cosmic
inflation yields density perturbations, the impact of which can be
observed in the measurement of CMB temperature
anisotropies~\cite{2dFGRS:2001csf, WMAP:2003syu, SDSS:2003tbn,
SDSS:2003eyi}. Moreover, to produce a long enough inflationary
era, it is essential to impose the (so-called) slow-roll
conditions on the inflaton field, under which the kinetic term of
the inflaton field can usually be ignored. On the other hand, a
wide range of potentials, that can be capable of describing an
inflationary era, have carefully been investigated and constrained
via the measurements of CMB anisotropies~\cite{Hossain:2014coa,
Martin:2013nzq, Geng:2015fla, Martin:2015dha, Huang:2015cke}.

On the other side, although GR has so far provided the most
accurate predictions for describing cosmological
phenomena~\cite{Will:2005va}, it is undesirable to justify the
dark sector (i.e., dark matter and dark energy) that its effects
on the dynamics of the Universe is well consistent with the
observational data~\cite{Ishak2019}. The need to overcome such
shortcomings motivated the investigation of alternative theories
of gravity, see
Refs.~\cite{Farhoudi2006,Felice,Sotiriou2010,Nojiri2011,capozziello2011,clifton,Farajollahi2012,Shabani2014,
Joyce2015,Bueno2016,zare1,khosravi2016, Nojiri:2017ncd,
quiros2019,Mishra2020} and references therein. In this regard,
various modified theories of gravity have also been considered for
inflationary cosmology in the hope that their predictions can
better justify the observational data~\cite{Myrzakulov:2015qaa,
DeLaurentis:2015fea,Sebastiani:2016ras,
Tirandari:2017nzy,saba,RSFM, Chakraborty:2018scm,Bernardo,
Kausar2019,
Bhattacharjee:2020jsf,Jalalzadel-et-al,Gamonal:2020itt,TQ.Do,baffou,faraji,Bhattacharjee,
Chen:2022dyq, Zhang:2021ppy}.

Another approach to structuring a more general theory than GR is
to consider geometry beyond the Riemannian one. The first study in
this area was carried out by Weyl to achieve a geometric
interpretation for electromagnetism and a unified theory for
gravitation and
electromagnetism~\cite{Weyl1918,Weyl1919,Wheeler2018}. He assumed
that under the parallel transformation of a vector, in addition to
the direction of the vector, its length also changes. In this
generalization of geometry, Weyl introduced a compensating vector
field. In this idea, the connection of the system can be
decomposed into two connections. One of those connections
describes the vector length, and the other one, a Levi-Civita
connection, expresses the vector direction under the parallel
transportation. Both of these connections are torsionless within
the framework of the Weyl geometry. Weyl tried to interpret the
introduced vector field as the electromagnetic potential, however
his theory was~not successful and even was rejected by
himself~\cite{Weyl1921}.

However, Dirac tried to revive the Weyl geometry. Indeed, he
showed~\cite{Dirac} that the difference between the Weyl geometry
and the Riemannian one is in the expression of the partial
derivative. In other words, under the Weyl gauge transformation,
the Weyl space is reduced to a Riemannian one with a metric that
is conformally related to the original metric. Nevertheless, it
has been claimed~\cite{Hayashi} that the Weyl theory cannot
describe the electromagnetic interaction because the vector field
introduced by Weyl does~not couple to the spinor, unlike the
electromagnetic potential. Although the success of the Weyl theory
was rather limited from the physical point of view, it presented
some interesting points. Indeed, one of the most important
features of the Weyl geometry is providing the nonzero covariant
derivative of the metric tensor, which geometrically led to a
quantitative definition called nonmetricity.

In order to achieve a simpler geometrical formalism for gravity,
the teleparallel equivalent of GR was proposed that uses the
Weitzenb\"{o}ck connection with zero curvature and nonmetricity
tensors but nonzero torsion~\cite{weiz,TEGR}.

Another formalism, known as the symmetric teleparallel gravity,
has been constructed using a connection with zero curvature and
torsion tensors but with a nonmetricity tensor describing
gravitational interactions~\cite{STG}. The symmetric teleparallel
gravity was modified~\cite{Jimenez2018} into the coincident GR and
$f(Q)$ gravity, which achieves a specific class with a vanishing
connection in the so-called coincident gauge. One of the most
important points of this theory is that it deprives gravity of any
inertial character and separates inertial and gravitational
effects, which cannot be done in GR~\cite{telepalatini}. The
symmetric teleparallel gravity provides another geometrical
description of gravitation that is dynamically equivalent to GR.
This theory is described by the Einstein-Hilbert action in the
absence of boundary terms. Its construction naturally leads to
self-accelerating cosmological solutions in the early and
late-time Universe. The cosmological and astrophysical aspects of
this theory have been investigated in
Refs.~\cite{Jimenez2018,Harko2018,Lu2019,BeltranJimenez:2019tme,
DAmbrosio:2021pnd, DAmbrosio:2021zpm,Narawade}.

Another proposed theory as an alternative to GR is $f(Q, T)$
gravity~\cite{Xu:2019sbp}, wherein the gravitational action is
determined by an arbitrary function of the nonmetricity scalar,
$Q$, and the trace of the energy-momentum tensor, $T$. This theory
is in fact an extension of the symmetric teleparallel gravity and
$f(Q)$ gravity~\cite{Jimenez2018,Harko2018}. Till now, some
cosmological aspects of $f(Q, T)$ gravity have been studied, e.g.,
observational constraints of $f(Q, T)$
gravity~\cite{Arora:2020tuk}, cosmological implications of its
Weyl-type gravity~\cite{Xu:2020yeg, Gadbail:2021kgd,
Gadbail:2021fjf}, energy conditions in $f(Q, T)$
gravity~\cite{Arora:2020iva, Arora:2021jik, Arora:2020met},
Friedmann-Lema\^{\i}tre-Robertson-Walker (FLRW) cosmology in $f(Q,
T)$ gravity~\cite{Godani:2021mld}, transit cosmological models of
$f(Q, T)$ gravity~\cite{Pradhan2021}, and dynamical aspects and
cosmic acceleration in $f(Q,T)$ gravity~\cite{Pati:2021ach,
Agrawal:2021rur,2022Arora, Pati:2021zew}. Also, a further
extension of $f(Q,T)$ gravity has been performed to squared
symmetric teleparallel gravity $f(Q,
T_{\mu\nu}T^{\mu\nu})$~\cite{Rudra}.

In this work, we intend to investigate the cosmological inflation
within the context of linear $f(Q, T)$ gravity. In this respect,
the outline of the work is as follows. In Sec.~II, we concisely
introduce the context of cosmological inflation. Then, in
Sec.~III, we briefly review the theoretical framework of $f(Q, T)$
gravity. Accordingly, in Sec.~IV, we calculate the cosmological
inflation within the context of such a gravity and hereupon,
discuss the relevant results. Moreover, in Sec.~V, we calculate
the cosmological inflation for a linear functional form of $f(Q,
T)$ gravity in the presence of a scalar field, and then, specify
inflationary observables by considering three different
inflationary potentials. Finally, in Sec.~VI, we scrutinize the
results and summarize the findings. Also, we provide an overview
of some prerequisites related to the $f(Q, T)$ gravity in an
appendix.

\section{Cosmological Inflation}
The simplest model of inflation is based on GR and an isotropic
and homogeneous scalar field called the inflaton. The dynamics of
such an inflaton field can be defined via the action
\begin{equation}\label{inflation}
S=\int \sqrt{-g}\left(\dfrac{R}{2\kappa}+L_{\rm
m}^{[\phi]}\right){\rm d}^4x,
\end{equation}
where $g$ is the determinant of the metric, $R$ is the Ricci
scalar, $\kappa=8\pi=1/{\rm M^2_{Pl}}$ (in the natural units
$\hbar=1=c$ and $G=1$) and the lower case Greek indices run from
zero to three. Also, $L_{\rm m}^{[\phi]}$ is the Lagrangian of the
inflaton field defined as
\begin{equation}\label{inflatonlagrangy}
L_{\rm
m}^{[\phi]}=-\dfrac{1}{2}g^{\mu\nu}\partial_{\mu}\phi\partial_{\nu}\phi-V(\phi),
\end{equation}
where $\phi\equiv\phi(t)$ is the scalar field and $V(\phi)$ is the
potential of the scalar field that depends on one or more free
parameters~\cite{lythandriotto}. During the inflationary era, the
inflaton slowly rolls down along its relatively flat potential.

The variation of action~\eqref{inflation} with respect to the
metric gives
\begin{equation}\label{enesteineq}
R_{\mu \nu}-\dfrac{1}{2}R g_{\mu\nu}=\kappa T_{\mu\nu},
\end{equation}
where $T_{\mu\nu}$ is the energy-momentum tensor that, generally,
defined as
\begin{equation}\label{tmunu}
T_{\mu\nu}\equiv -\dfrac{2}{\sqrt{-g}}\frac{\delta(\sqrt{-g}L_{\rm
m})}{\delta g^{\mu\nu}}=g_{\mu\nu}L_{\rm m}-2\dfrac{\partial
L_{\rm m}}{\partial g^{\mu\nu}}.
\end{equation}
By substituting relation~\eqref{inflatonlagrangy} into
relation~\eqref{tmunu}, it gives
\begin{equation}\label{Tphii}
T_{\mu\nu}^{[\phi]}=\partial_\mu\phi\partial_\nu\phi-g_{\mu\nu}\left[\dfrac{1}{2}
\partial_\sigma\phi\partial^\sigma\phi+V(\phi)\right].
\end{equation}
We also assume the spatially flat FLRW spacetime
\begin{equation}\label{frwmetric}
ds^{2}=-dt^{2}+a^{2}(t)(dx ^{2}+ dy ^{2}+dz^{2}),
\end{equation}
where $a(t)$ is the cosmological scale factor as a function of the
cosmic time $t$ with the present time Universe $t_{0}$ as
$a(t_{0})=1$. The energy-momentum tensor of the inflaton field can
be declared as a perfect fluid with a linear barotropic equation
of state as $p^{[\phi]}=w\rho^{[\phi]}$, in which the energy
density, $\rho^{[\phi]}$, and the pressure density, $p^{[\phi]}$,
are
\begin{equation}\label{rho-p}
\rho^{[\phi]}=\dfrac{1}{2}\dot{\phi}^2+V(\phi)\qquad{\rm
and}\qquad p^{[\phi]}=\dfrac{1}{2}\dot{\phi}^2-V(\phi),
\end{equation}
where dot represents the derivative with respect to the cosmic
time. The dimensionless related parameter of the equation of state
also is
\begin{equation}\label{w}
w^{[\phi]}=\dfrac{p^{[\phi]}}{\rho^{[\phi]}}=\dfrac{\dot{\phi}^2-2V}{\dot{\phi}^2+2V}.
\end{equation}
Now, under these considerations and by using
Eqs.~\eqref{enesteineq}, \eqref{Tphii}, \eqref{frwmetric} and
\eqref{rho-p}, we easily obtain the Friedmann equations as
\begin{align}
\label{firsth2}&H^2=\dfrac{\kappa}{3}\left[\dfrac{\dot{\phi}^2}{2}+V(\phi)\right],\\
&H^2+\dot{H}=-\dfrac{\kappa}{3}\left[\dot{\phi}^2-V(\phi)\right],\\
\label{firsthdot}&\dot{H}=-\dfrac{\kappa}{2}\dot{\phi}^2,
\end{align}
where $H(t)\equiv\dot{a}/a$ is the Hubble parameter. To achieve
the time evolution of the scalar field, by taking the time
derivative of Eq.~\eqref{firsth2} and substituting
Eq.~\eqref{firsthdot} into the resulted relation, we obtain the
Klein-Gordon equation as
\begin{equation}\label{firstkg}
\ddot{\phi}+3H\dot{\phi}+V^{\prime}=0,
\end{equation}
where the prime denotes derivative with respect to the scalar
field.

Inflation is an accelerated expansion era in the early Universe,
wherein the comoving Hubble horizon shrinks in time, i.e.,
\begin{equation}
\dfrac{d(aH)^{-1}}{dt}=-\dfrac{\ddot{a}}{\dot{a}^2}=-\dfrac{1}{a}\left(1-\epsilon_1\right)<0.
\end{equation}
In this relation, $\epsilon_1<1$ is the first slow-roll parameter,
which is defined as~\cite{liddle1994, mokhanov}
\begin{equation}\label{eps1}
{\rm\epsilon_1}(t)\equiv-\dfrac{\dot{H}}{H^2}.
\end{equation}
Moreover, there are several possible sets of such slow-roll
parameters that are useful to define in terms of the e-folding
number, $N$, namely~\cite{ency}
\begin{equation}\label{hubbleflow}
{\rm\epsilon_{n+1}}(t)\equiv\dfrac{\rm{d\ln|\epsilon_n(t)|}}{\rm{dN}},
\end{equation}
where $n\geq 0$ is an integer and ${\rm\epsilon_0}(t)\equiv H_{\rm
end}/H$. This relation is known as the Hubble flow parameters or
the Hubble slow-roll parameters. The e-folding number somehow
describes the rate of the expansion of inflation as a natural
logarithm of the scale factor~\cite{intcosminf, tasi lec}
\begin{equation}\label{Ndef}
N\equiv \ln(\dfrac{a_{\rm end}}{a})=\int_{t}^{t_{\rm end}} H\rm{dt},
\end{equation}
where the index `${\rm end}$' denotes the value of quantities at
the end of inflation. From definition~\eqref{hubbleflow}, the
second slow-roll parameter is
\begin{equation}\label{eps2}
{\rm\epsilon_2}=\dfrac{\dot{{\rm\epsilon_1}}}{H{\rm\epsilon_1}}=\dfrac{\ddot{H}}{\dot{H}H}-2\dfrac{\dot{H}}{H^2}.
\end{equation}
It is known~\cite{ency} that if the condition
$|{\rm\epsilon_n}|\ll 1$ is met, inflation will occur and will
continue long enough to solve the standard cosmological problems.
Also, inflation ends when the first slow-roll parameter reaches
the unit, i.e., ${\rm\epsilon_1}=1$.
 \vskip0.5cm
\noindent {\bf Potential Representation of Slow-Roll Parameters}

As mentioned, an inflationary scenario at the early stages of the
formation of the Universe can be described by the slow-roll
conditions. Hence, calculating the slow-roll parameters while
satisfying the slow-roll conditions can be considered as the first
step in inflationary calculations. These parameters can also be
written in terms of the inflationary potential used.

By substituting Eqs.~\eqref{firsth2} and \eqref{firsthdot} into
definition~\eqref{eps1}, one obtains the first slow-roll
parameter, usually denoted by $\epsilon$, in terms of scalar field
as
\begin{equation}
\label{1sr}{\rm \epsilon}\equiv \epsilon_1=\dfrac{3}{2}\dfrac{\dot{\phi}^2}{\dfrac{\dot{\phi}^2}{2}+V}.
\end{equation}
As a result, satisfying $\epsilon\ll 1$ leads to
\begin{equation}\label{1srr}
\dot{\phi}^2\ll |V|.
\end{equation}
By applying the above condition to relation~\eqref{1sr},
$\epsilon$ can be approximated as
\begin{equation}
\epsilon\approx\dfrac{3}{2}\dfrac{\dot{\phi}^2}{|V|}.
\end{equation}
Similar to the first slow-roll parameter, the usual second
slow-roll parameter can also be defined as
\begin{equation}\label{2sr}
\eta\equiv
2\epsilon-\dfrac{\epsilon_2}{2}=-\dfrac{\dot{H}}{H^2}-\dfrac{\ddot{H}}{2H\dot{H}}\approx-\dfrac{|\ddot{\phi}|}{H|\dot{\phi}|},
\end{equation}
where, in the last approximation, relation~\eqref{1srr} and
Eqs.~\eqref{firsth2}-\eqref{firsthdot} have been used.
Furthermore, the condition $|\eta|\ll 1$ yields
\begin{equation}\label{2srr}
|\ddot{\phi}|\ll H|\dot{\phi}|,
\end{equation}
which, together with condition~\eqref{1srr}, are known as the
slow-roll conditions. The slow-roll conditions manage the dynamics
of inflationary scenarios in such a way that it will lead to
inflation if these conditions can be fulfilled, and when those are
violated, inflation will end. Accordingly, by using the slow-roll
conditions, Eqs.~\eqref{firsth2} and \eqref{firstkg} can also be
approximated to
\begin{align}\label{approx11}
&H^2\approx\dfrac{\kappa}{3}V(\phi),\\
\label{1phidot}
&\dot{\phi}\approx-\dfrac{V^{\prime}}{3H}.
\end{align}
Now, via Eqs. \eqref{approx11} and \eqref{1phidot}, the slow-roll
parameters can be written in terms of the inflationary potential
and its derivatives as
\begin{equation}\label{epsv}
\epsilon\approx\dfrac{1}{2\kappa}\left(\dfrac{V^{\prime}}{V}\right)^2,
\end{equation}
and~\cite{liddle1994, weinberg, myrzak}
\begin{equation}\label{etav}
\eta\approx \dfrac{1}{\kappa}\dfrac{V^{\prime\prime}}{V}.
\end{equation}
Relations \eqref{epsv} and \eqref{etav} are known as the potential
slow-roll parameters, which are distinct from the Hubble slow-roll
parameters. Using the slow-roll approximations and the equation of
state as $w^{[\phi]}\approx -1$, the e-folding number,
relation~\eqref{Ndef}, can be calculated in terms of the
inflationary potential as
\begin{equation}\label{nGR}
N=\int_{\phi}^{\phi_{\rm end}}\dfrac{H}{\dot{\phi}}{\rm d}\phi=\kappa\int_{\phi_{\rm end}}^{\phi}\dfrac{V}{V^{\prime}}{\rm d}\phi.
\end{equation}
Up to now, we have introduced the slow-roll parameters in both
Hubble and potential representations within the context of the
cosmological inflation scenario. In the following, we will briefly
review the theoretical framework of $f(Q, T)$ gravity and its
cosmological implications.

\section{The Framework of $f(Q, T)$ Gravity}
To avoid any digressions, we review some prerequisites related to
the $f(Q,T)$ gravity in an appendix.

Following Ref.~\cite{Xu:2019sbp}, we consider the $f(Q,T)$ gravity
action
\begin{equation}\label{fqtaction}
S=\int \sqrt{-g}\left[\dfrac{f(Q, T)}{2\kappa}+L_{\rm
m}\right]{\rm d}^{4}x,
\end{equation}
where in addition to the metric, this action also varies with
respect to the connection, which is assumed to be torsionless with
a zero Riemann tensor, while accepting nonmetricity.

The variation of action \eqref{fqtaction} with respect to the
metric leads to the modified field equations
 \begin{equation}
\begin{array}{l}
 \kappa\, T_{\mu\nu}=-\dfrac{2}{\sqrt{-g}}\bigtriangledown_{\alpha}
 \left(f_{Q}\sqrt{-g}P^{\alpha}{}_{\mu\nu}\right)-\dfrac{f\,g_{\mu\nu}}{2}\\ \\
\kern 1.4pc +f_{T}\left(T_{\mu\nu}+\Theta_{\mu\nu}\right) -f_{Q}\left(P_{\mu\alpha\beta}
Q_{\nu}{}^{\alpha\beta}-2Q^{\alpha\beta}{}_{\mu}P_{\alpha\beta\nu}\right),
\end{array}
\end{equation}
where
\begin{equation}
\Theta_{\mu\nu}\equiv g^{\alpha\beta}\dfrac{\delta
T_{\alpha\beta}} {\delta g^{\mu\nu}},~~ f_{Q}\equiv\dfrac{\partial
f(Q,T)}{\partial Q},~~ f_{T}\equiv\dfrac{\partial f(Q,T)}{\partial
T}.
\end{equation}
Furthermore, $P^{\alpha}{}_{\mu\nu}$ is the superpotential of the
model, which is defined as
\begin{equation}
P^{\alpha}{}_{\mu\nu}\equiv -\dfrac{1}{2}L^{\alpha}{}_{\mu\nu}+\dfrac{1}{4}
\left(Q^{\alpha}-\tilde{Q}^{\alpha}\right)g_{\mu\nu}-\dfrac{1}{4}\delta^{\alpha}_{(\mu}Q_{\nu)}.
\end{equation}
In this relation, each of ${Q}^{\alpha}$ and $\tilde{Q}^{\alpha}$
is the trace of the nonmetricity tensor defined as
\begin{equation}
{Q}_{\alpha}\equiv Q_{\alpha}{}^{\mu}{}_{\mu}\qquad{\rm
and}\qquad\tilde{Q}_{\alpha}\equiv Q^{\mu}{}_{\alpha\mu}.
\end{equation}

The variation of action \eqref{fqtaction} with respect to the
connection can be performed while imposing two constraints
$R^{\alpha}{}_{\beta\mu\nu}=0$ and $\tau^{\alpha}{}_{\mu\nu}=0$,
and using the Lagrange multiplier method. Consequently the
corresponding field equations are
\begin{equation}
\bigtriangledown_{\mu}\bigtriangledown_{\nu}\(2\sqrt{-g}f_{Q}P^{\mu\nu}{}_{\alpha}+\kappa\,
H_{\alpha}{}^{\mu\nu}\)=0,
\end{equation}
where $H_{\alpha}{}^{\mu\nu}$ is the hyper-momentum tensor density defined as
\begin{equation}
H_{\alpha}{}^{\mu\nu}\equiv\dfrac{f_{T}\sqrt{-g}}{2\kappa}\dfrac{\delta\,
T}{\delta\Gamma^{\alpha}{}_{\mu\nu}}+\dfrac{\delta\(\sqrt{-g}\,{\rm
L_m}\)}{\delta \Gamma^{\alpha}{}_{\mu\nu}}.
\end{equation}

We also assume the spatially flat FLRW spacetime,
metric~\eqref{frwmetric}, and consider the energy-momentum tensor
as a perfect fluid, i.e.,
\begin{equation}
T_{\mu\nu}=\left(\rho+p\right)u_\mu u_\nu+p\,g_{\mu\nu},
\end{equation}
where $\rho$, $p$ and $u^\mu$ are the energy density, the pressure
density, and the four-vector velocity of the perfect fluid,
respectively. In an adapted coordinate system, i.e. the Cartesian
coordinates in the spatial variables of the
line-element~\eqref{frwmetric} in which the connection is zero
(the coincident gauge), one obtains the nonmetricity scalar to be
$Q=6H^2$. Hence the modified Friedmann equations are
\begin{align}
\label{rho}&\kappa\,\rho=\dfrac{f}{2}-6FH^2-\dfrac{2\tilde{G}}{1+\tilde{G}}\left(\dot{F}H+F\dot{H}\right),\\
\label{p}&\kappa\,
p=-\dfrac{f}{2}+6FH^2+2\left(\dot{F}H+F\dot{H}\right),
\end{align}
where $F\equiv f_{Q}$ and $\kappa\tilde{G}\equiv f_{T}$ have been
defined for simplicity. Under these situations, one obtains the
evolution of the Hubble parameter as
\begin{equation}
\label{Hdot}\dot{H}+\dfrac{\dot{F}}{F}H=\dfrac{\kappa}{2F}\left(1+\tilde{G}\right)\left(\rho+p\right),
\end{equation}
where $F\neq 0$ has been assumed\rlap.\footnote{In the case $F=0$,
Eqs.~\eqref{rho} and \eqref{p} give $w=-1$.}\
 By introducing the effective pressure and energy densities, one
can rewrite the modified Friedmann equations as
\begin{align}
\label{rhoeff}&3H^2=\kappa\rho^{[\rm eff]}=\dfrac{f}{4F}-\dfrac{\kappa}{2F}\left[\left(1+\tilde{G}\right)\rho+\tilde{G}p\right],\\
&2\dot{H}+3H^2=-\kappa p^{[\rm eff]}=\dfrac{f}{4F}-\dfrac{2\dot{F}H}{F}\nonumber\\
\label{peff}&~~~~~~~~~~~~~~~~~+\dfrac{\kappa}{2F}\left[\left(1+\tilde{G}\right)\rho+\left(2+\tilde{G}\right)p\right].
\end{align}
Also, taking the time derivative of Eq.~\eqref{rhoeff} while using
Eqs.~\eqref{rhoeff} and \eqref{peff} yields the convenient
continuity equation
\begin{equation}
\dot{\rho}^{[\rm eff]}+3H\left(\rho^{[\rm eff]}+p^{[\rm eff]}\right)=0.
\end{equation}
Moreover, by substituting Eq.~\eqref{Hdot} into Eq.~\eqref{rho}
while using the linear barotropic equation of state, we obtain the
evolution of energy density to be
\begin{equation}\label{rhooo}
\rho=\dfrac{f-12H^2F}{2\kappa\left[(1+w)\tilde{G}+1\right]}.
\end{equation}

\section{Modeling an Inflationary Scenario in $f(Q, T)$ Gravity}
Inflation can be realized by a scalar field called inflaton or
from modification of gravity. In this section, we analyze the
slow-roll inflation within the framework of a special model of the
$f(Q,T)$ gravity. That is, we consider a linear functional form of
$f(Q,T)$ gravity, i.e. $f(Q,T)=\alpha\, Q+\beta\, T$, with
$\alpha=F\neq 0$ and $\beta$ as free constant parameters of the
model. The model obviously reduces to $f(Q)$ gravity for $\beta=0$
that has been considered in, e.g., Ref.~\cite{Hassan2021}, and the
cosmological evolution of $f(Q,T)$ gravity has been studied in
Ref.~\cite{Xu:2019sbp}.

By taking a linear barotropic equation of state and $T=-\rho+3p$,
Eqs.~\eqref{Hdot}, \eqref{rhoeff}, \eqref{peff} and \eqref{rhooo}
reduce to
\begin{align}
\label{Hdoooot}&\dot{H}=\dfrac{\left(\kappa+\beta\right)\rho\left(1+w\right)}{2\alpha},\\
\label{frw11}&3H^2=\kappa\rho^{[\rm eff]}=\dfrac{\left[(w-3)\beta-2\kappa\right]\rho}{2\alpha},\\
\label{frw22}&2\dot{H}+3H^2=-\kappa p^{[{\rm eff}]}=\dfrac{\rho\left[\left(3w-1\right)\beta+2w\kappa\right]}{2\alpha},\\
\label{rhhho}&\rho=\dfrac{-6\alpha
H^2}{\beta\left(3-w\right)+2\kappa},
\end{align}
where
\begin{equation}\label{constraints}
(1+w)\beta+\kappa\neq 0\qquad {\rm and}\qquad
(w-3)\beta-2\kappa\neq 0
\end{equation}
have been assumed\rlap.\footnote{These conditions obviously
dictate that for $\forall\beta$, except $\beta\neq 0$, one has
$w\neq -1-\kappa/\beta$ and $w\neq 3+2\kappa/\beta$. As a few
examples, these conditions include ($w=-1$ with $\beta\neq
-\kappa/2$), ($w=0$ with $\beta\neq -2\kappa/3$ and $\beta\neq
-\kappa$), ($w=1$ with $\beta\neq -\kappa/2$ and $\beta\neq
-\kappa$), ($\beta= -\kappa/2$ with $w\neq \pm 1$) and ($\beta=
-\kappa$ with $w\neq 0$ and $w\neq 1$).}\
 In addition, in the case of $\beta =-\kappa$ and/or $w=-1$,
Eq.~\eqref{Hdoooot} gives $\dot{H}=0$, and hence $H$ (and in turn,
$\rho$) is constant. These cases lead either to a de~Sitter
expansion for nonzero $H$ or to the Minkowski metric for $H=0$.
Therefore, we also exclude these values in the numerical
calculations of inflationary observables.

Note that, contrary to what has been claimed in
Ref.~\cite{Xu:2019sbp}, no consistency condition for the parameter
$\beta$ can be obtained from Eqs.~\eqref{Hdoooot}, \eqref{frw11}
and \eqref{frw22}\rlap.\footnote{Indeed by different combinations
of these equations, one simply obtains exact relations
$(45)-(44)=(43)$, $(45)-2(43)=(44)$ and $2(43)+(44)=(45)$ (as
expected, the Friedmann equations are~not independent).}\
 Hence, the evolution of the linear
model of $f(Q,T)$ gravity is~not fixed on a de~Sitter expansion.

In continuation, Eqs.~\eqref{frw11} and \eqref{frw22} lead to an
effective parameter of the equation of state for this model as
\begin{equation}\label{effw}
w^{[{\rm eff}]}=\dfrac{p^{[{\rm eff}]}}{\rho^{[{\rm eff}]}}=-1-\dfrac{2\left(\kappa+\beta
\right)\left(1+w\right)}{\left(w-3\right)\beta-2\kappa}.
\end{equation}
In this situation, different conditions can be considered as
follows.
\begin{itemize}
\item
The effective de~Sitter accelerated expansion (i.e., $w^{[{\rm
eff}]}=\,-1$) will be realized if one has either ($w=-1$ with
$\beta\neq -\kappa/2$) or ($\beta= -\kappa$ with $w\neq 1$).
However, we have already excluded $\beta =-\kappa$ and  $w=-1$
values.
\item
The effective phantom accelerated phase (i.e., $w^{[{\rm
eff}]}<-1$) will occur if one has ($\beta= 0$ with $w<-1$), or
($\beta > 0$ with either $w<-1$ or $w>3+2\kappa/\beta$), or
($\beta<-\kappa$ with either $w<-1$ or $w>3+2\kappa/\beta$), or
($-\kappa<\beta<-\kappa/2$ with $-1<w<3+2\kappa/\beta$), or
($-\kappa/2<\beta<0$ with $3+2\kappa/\beta<w<-1$).
\item
The effective quintessence accelerated evolution (i.e.,
$-1<w^{[{\rm eff}]}<-1/3$) will happen if one has ($\beta=0$ with
$-1<w<-1/3$, as expected), or ($w=0$ with $\beta<-\kappa$), or
($\beta<-\kappa$ with $-1<w<-\kappa/(3\kappa+4\beta)$ [note that,
this value includes $w=0$ and for $\beta\rightarrow -\kappa$, its
upper limit is $w<1$]), or ($-\kappa<\beta<-3\kappa/4$ with either
$w>-\kappa/(3\kappa+4\beta)$ or $w<-1$), or ($\beta=-3\kappa/4$
with $w<-1$), or ($-3\kappa/4<\beta<-2\kappa/3$ with
$-\kappa/(3\kappa+4\beta)<w<-1)$), or ($\beta=-2\kappa/3$ with
$-3<w<-1$), or ($-2\kappa/3<\beta<-\kappa/2$ with
$-\kappa/(3\kappa+4\beta)<w<-1$), or ($-\kappa/2<\beta<0$ with
$-1<w<-\kappa/(3\kappa+4\beta)$), or ($\beta>0$ with
$-1<w<-\kappa/(3\kappa+4\beta)$).
\end{itemize}

On the other hand, by substituting Eq.~\eqref{rhhho} into
Eq.~\eqref{Hdoooot} the following differential equation can be
obtained
\begin{equation}\label{hdothw}
\dot{H}-3A(w,\beta)H^2=0,
\end{equation}
where
\begin{equation}
A(w,\beta)\equiv\dfrac{\left(w+1\right)\left(\kappa+\beta\right)}{\left(w-3\right)\beta-2\kappa}.
\end{equation}
Eq.~\eqref{hdothw} is a differential equation for the Hubble
parameter that gives
\begin{equation}\label{h}
 H=\dfrac{\beta\left(w-3\right)-2\kappa}{\beta\left[w
\left(-3t+C\right)-3t-3C\right]-\kappa\left(3tw+3t+2C \right)},
\end{equation}
where $C$ is an integration constant. However, the time
derivatives of the Hubble parameter can also be extracted from the
above relation as
\begin{align}
\label{hhdot}&\dot{H}\!=\!\dfrac{3\left(w+1\right)\left(\kappa+\beta\right)
\left[\beta\left(w-3\right)-2\kappa\right]}{\big\lbrace\beta\left[w
\left(-3t+C\right)-3t-3C\right]-\kappa\left(3tw+3t+2C
\right)\!\big\rbrace^2},\\\label{hhddot}&\ddot{H}\!=\!\dfrac{18\left(w+1
\right)^2\left(\kappa+\beta\right)^2\left[\beta\left(w-3\right)-2\kappa
\right]}{\big\lbrace\beta\left[w\left(-3t+C\right)-3t-3C\right]-\kappa
\left(3tw+3t+2C\right)\!\big\rbrace^3}.
\end{align}

The main purpose of gravitational theories in properly describing
cosmological inflation is to provide consistent predictions with
the observational data. For this purpose, one has to examine the
theoretical predictions of the proposed gravitational models for
the inflationary observables that are as follows~\cite{ency,
liddlebook,higer,Lythandliddle}.
\begin{itemize}
\item The scalar spectral index:
\begin{equation}
\label{ns}n_{\rm S}=1+\dfrac{{\rm d}\ln(\Delta_{\rm S}^2)}{{\rm
d}\ln {\rm k}}=1-6\epsilon+2\eta.
\end{equation}
\item The tensor spectral index:
\begin{equation}
\label{nt}n_{\rm T}=\dfrac{{\rm d}\ln(\Delta_{\rm T}^2)}{{\rm
d}\ln{\rm k}}=-2\epsilon.
\end{equation}
\item The tensor-to-scalar ratio:
\begin{equation}
\label{r}r=\dfrac{\Delta_{\rm T}^2({\rm k})}{\Delta_{\rm S}^2({\rm
k})}=16\epsilon.
\end{equation}
\end{itemize}
In relations~\eqref{ns}, \eqref{nt} and \eqref{r}, $\Delta_{\rm
S}$ and $\Delta_{\rm T}$ are respectively the dimensionless power
spectrum for scalar perturbations and tensor perturbations, and
${\rm k}=aH$. By substituting Eqs.~\eqref{h}, \eqref{hhdot} and
\eqref{hhddot} into Eqs.~\eqref{eps1}, \eqref{eps2} and
\eqref{2sr}, the slow-roll parameters for the proposed
gravitational model are
\begin{equation}\label{srparameter}
{\rm\epsilon}=\dfrac{-3\left(w+1\right)\left(\kappa+\beta\right)}{\beta\left(w-3\right)-2\kappa},\quad
{\rm\epsilon_2}=0\quad{\rm and}\quad \eta=2\epsilon.
\end{equation}

\begin{figure*}[t!]
\centering
\subfigure[]{
\includegraphics[width=0.31\textwidth]{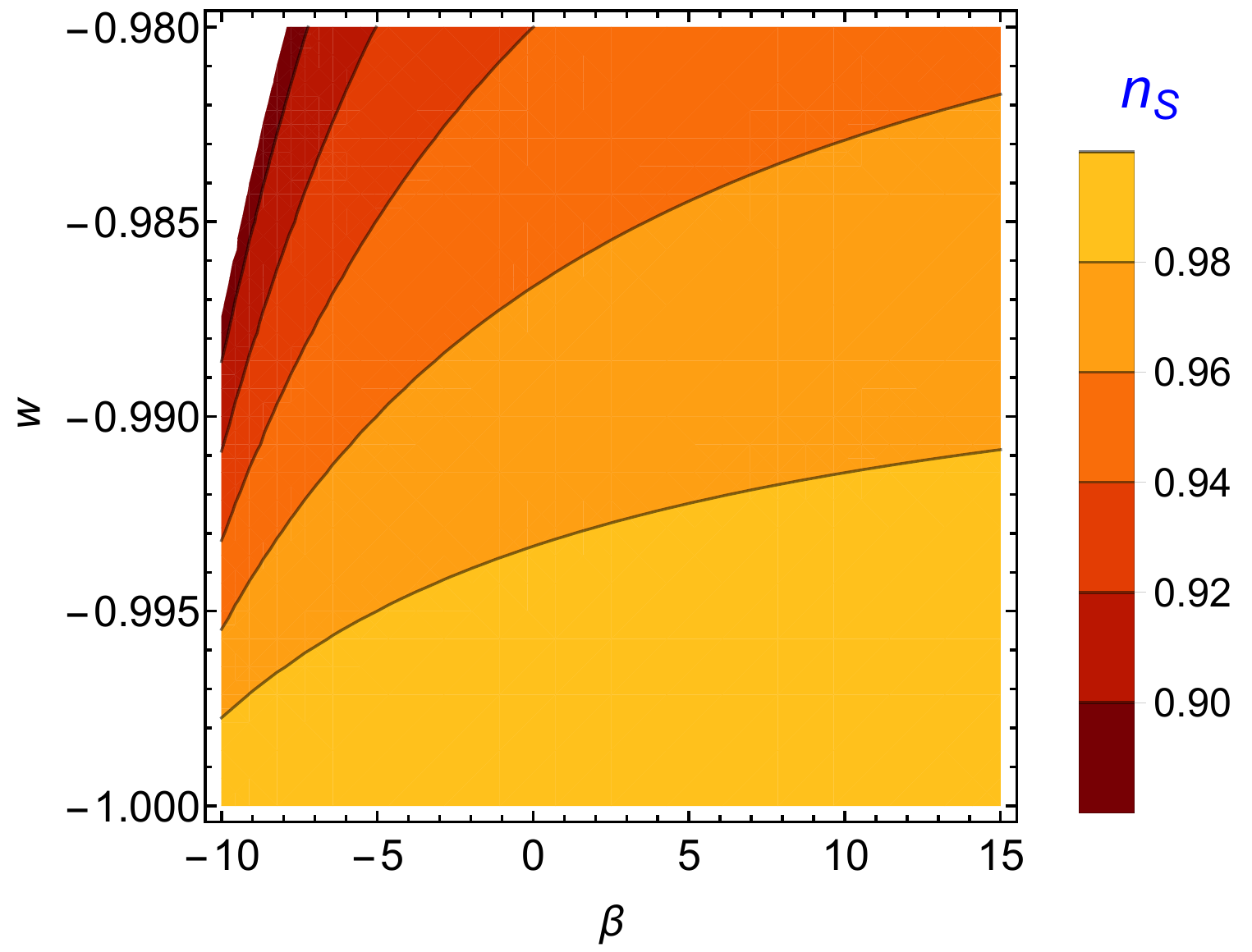}
\label{aa1}
}
\subfigure[]
{
\includegraphics[width=0.32\textwidth]{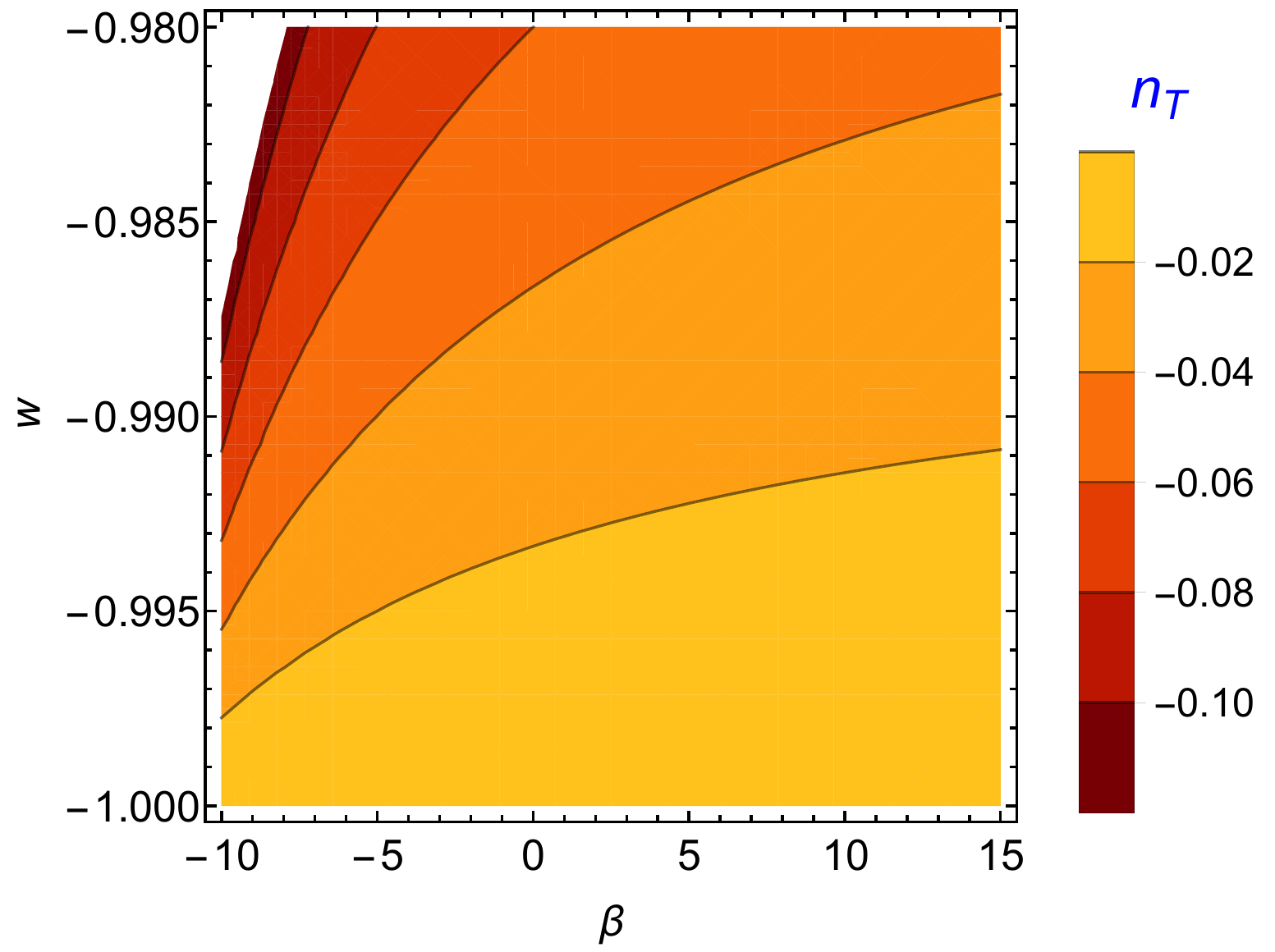}
\label{bb1}
}
\subfigure[]
{
\includegraphics[width=0.31\textwidth]{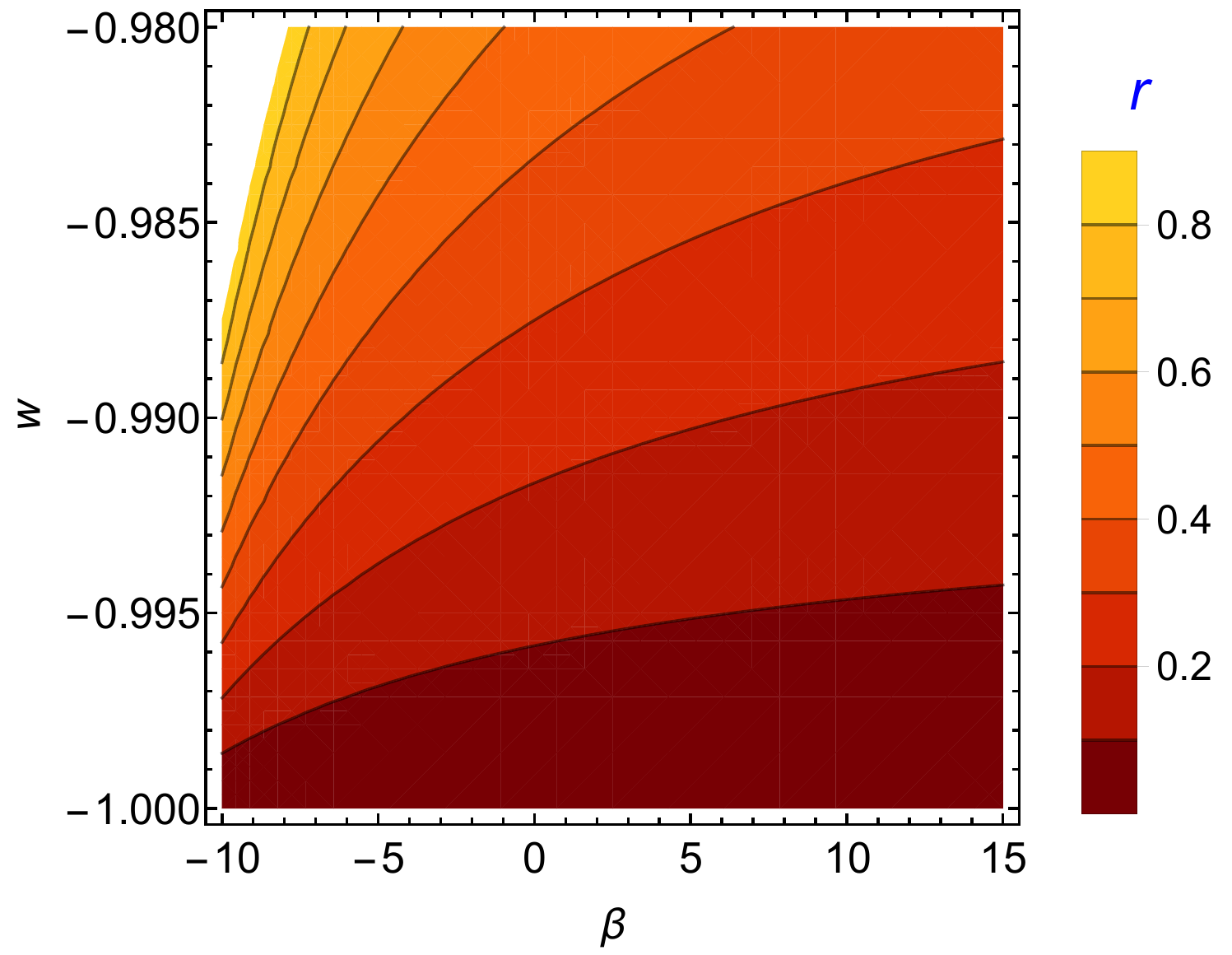}
\label{cc1} } \caption{\label{fig1} [color online] Inflationary
observables as functions of the dimensionless parameter of the
equation of state $w$ and the free parameter $\beta$, wherein (a)
shows the scalar spectral index, $n_{\rm S}$, (b) indicates the
tensor spectral index, $n_{\rm T}$, and (c) demonstrates the
tensor-to-scalar ratio, $r$.}
\end{figure*}

\begin{table*}[t]
\caption{Inflationary observables for different values of $w$ and
$\beta$.} \centering
\begin{tabular}{c | c | c | c | c}
\hline
\hline
$w$ & $\beta$ & $n_{\rm S}$ & $r$ & $n_{\rm T}$ \\ [0.5ex]
\hline
-0.984 & 14.1 & 0.96461 & 0.28305 & -0.03538 \\
-0.986 & 5.4 & 0.96427 & 0.28578 & -0.03572 \\
-0.988 & 0.3 & 0.96441 & 0.28466 & -0.03558 \\
-0.990 & -3.5 & 0.96423 & 0.28608 & -0.03576 \\
-0.992 & -6.2 & 0.96437 & 0.28502 & -0.03562 \\
-0.994 & -8.3 & 0.96456 & 0.28345 & -0.03543 \\
-0.996 & -0.29 & 0.98785 & 0.09713 & -0.01214 \\
-0.998 & -8.5 & 0.98773 & 0.09814 & -0.01226 \\
\hline
\hline
\end{tabular}
\label{table1}
\end{table*}

Now, let us calculate the inflationary observables in the $f(Q,
T)=\alpha\, Q + \beta\, T$ gravity and compare the results with
the observational data. In this regard, the latest constraints
from the Planck collaboration on the scalar spectral index and the
tensor-to-scalar ratio are~\cite{Planck:2018jri}
\begin{eqnarray}\label{planckdata}
&n_{\rm S}= 0.9649\pm 0.0042\quad {\rm at}\ 68 \%\, {\rm CL},\cr
 &\cr
 & r<0.10\quad {\rm at}\ 95 \%\, {\rm CL}.
\end{eqnarray}
However, by the joint Planck, BK15 and BAO data, further
constraint tightened the upper limit on r to be
\begin{equation}\label{planckdata2}
r<0.056\quad{\rm at}\ 95 \%\, {\rm CL}.
\end{equation}
On the other hand, using the slow-roll parameters, according to
relations~\eqref{ns}, \eqref{nt} and \eqref{r} and
\eqref{srparameter}, the inflationary observables for the proposed
gravitational model are obtained to be
\begin{align}
&n_{\rm S}=\dfrac{\left(7w+3\right)\beta+\left(4+6w\right)\kappa}{\beta\left(w-3\right)-2\kappa},\\
&n_{\rm T}=\dfrac{6\left(w+1\right)\left(\kappa+\beta\right)}{\beta\left(w-3\right)-2\kappa},\\
&r=\dfrac{-48\left(w+1\right)\left(\kappa+\beta\right)}{\beta\left(w-3\right)-2\kappa}.
\end{align}
As is clear from the above relations, the inflationary observables
do~not depend on the free parameter $\alpha$ and the e-folding
number. Instead, the dependence of these observations is on the
free parameter $\beta$ and the dimensionless parameter of the
equation of state $w$.

In order to realize the robustness of the model in the case of
cosmological inflation, the comparison of the free parameters of
the model with the latest observational data have been
investigated. The numerical results for the scalar spectral index,
the tensor spectral index, and the tensor-to-scalar ratio as
functions of $w$ and $\beta$ have been shown in Fig.~\ref{fig1} in
the range of $[-1.000,-0.980]$ and $[-10,15]$, respectively. It
can be inferred from the contour plots that the values of both $w$
and $\beta$ parameters affect on the values of the inflationary
observables. Also, the consistent values of inflationary
observables for some values of parameters $w$ and $\beta$ have
been given in Table~\ref{table1}. As is clear, the theoretical
predictions of the proposed gravitational model for some values of
$w$ and $\beta$ for $n_{\rm S}$ and $r$ are in agreement with the
observational data. However, for the values ($w=-1$ with
$\beta\neq -\kappa/2$) and/or ($\beta=-\kappa$ with $w\neq 1$),
one obtains a de~Sitter expansion, which indicates that inflation
does~not occur (besides, $\rho$ is constant in these cases). In
addition, for the values ($w=-1$ with $\beta = -\kappa/2$) and/or
($\beta=-\kappa$ with $w= 1$), which do~not satisfy the second
condition~\eqref{constraints}, Eq.~\eqref{frw11} implies $H=0$,
and in turn $a(t)$ to be constant, and thus the solution to the
field equations is the Minkowski metric with no inflation.

\section{Slow-Roll Inflation in $f(Q,T)$ Gravity with Scalar Field}
In this section, we are going to analyze the cosmological
inflation within the context of the linear $f(Q,T)$ gravity in the
presence of a scalar field as the matter Lagrangian, i.e.
relation~\eqref{inflatonlagrangy}, and as a perfect fluid with the
associated linear barotropic equation of state that is just a
function of the cosmic time. Hence by relation~\eqref{Tphii}, the
trace of the energy-momentum tensor is
\begin{equation}
T^{[\phi]}=\dot{\phi}^2-4V.
\end{equation}
Also, by substituting relations~\eqref{rho-p} into
Eqs.~\eqref{rhoeff} and \eqref{peff}, we obtain the components of
the effective energy-momentum tensor as
\begin{align}
\label{rho11}&\rho^{[\rm eff]}=-\dfrac{\left(\kappa+\beta\right)\dot{\phi}^2+2\,V(\phi)\left(\kappa+2\beta\right)}{2\kappa\alpha},\\
\label{p11}&p^{[\rm
eff]}=-\dfrac{\left(\kappa+\beta\right)\dot{\phi}^2-2\,V(\phi)\left(\kappa+2\beta\right)}{2\kappa\alpha}.
\end{align}
Accordingly, the effective dimensionless parameter of the equation
of state obviously is
\begin{equation}\label{weff}
w^{[\rm
eff]}=\dfrac{\left(\kappa+\beta\right)\dot{\phi}^2-2\,V(\phi)\left(\kappa+2\beta
\right)}{\left(\kappa+\beta\right)\dot{\phi}^2+2\,V(\phi)\left(\kappa+2\beta\right)}.
\end{equation}

Moreover, by considering Eqs.~\eqref{rho11} and \eqref{p11}, the
modified Friedmann equations~\eqref{rhoeff} and \eqref{peff} can
be translated into the formulas
\begin{align}
\label{HH2}&\dfrac{3}{2}H^2=-\dfrac{\left(\kappa+\beta\right)\dot{\phi}^2+2\,V(\phi)\left(\kappa+2\beta\right)}{4\alpha},\\
&2\dot{H}+\dfrac{3}{2}H^2=\dfrac{3\left(\kappa+\beta\right)\dot{\phi}^2-2\,V(\phi)\left(\kappa+2\beta\right)}{4\alpha}.
\end{align}
From these relations, $\dot{H}$ reads
\begin{equation}\label{HHdot}
\dot{H}=\dfrac{\dot{\phi}^2}{2\alpha}\left(\kappa+\beta\right),
\end{equation}
therefore we exclude $\beta=-\kappa$ again. By taking the time
derivative of Eq.~\eqref{HH2} and substituting Eq.~\eqref{HHdot}
into the result, we obtain the modified Klein-Gordon equation as
\begin{equation}\label{kgp}
\ddot{\phi}\left(\kappa+\beta\right)+3H\dot{\phi}\left(\kappa+\beta\right)
+V^{\prime}\left(\kappa+2\beta\right)=0.
\end{equation}

Now, let us calculate the extended slow-roll parameters within the
context of the linear $f(Q,T)$ gravity. For this purpose, one has
to insert Eqs.~\eqref{HH2} and \eqref{HHdot} into
Eq.~\eqref{eps1}. Accordingly, the first extended slow-roll
parameter is obtained as
\begin{equation}\label{epssss}
{\rm\epsilon}=\dfrac{3\left(\kappa+\beta\right)\dot{\phi}^2}{\left(\kappa+\beta
\right)\dot{\phi}^2+2\,V\left(\kappa+2\beta\right)}.
\end{equation}
As inflation will occur if ${\rm \epsilon}\ll 1$, hence the
slow-roll condition~\eqref{1srr} within the context of the linear
$f(Q,T)$ gravity in this situation will be modified as
\begin{equation}\label{exsr}
\left(\kappa+\beta\right)\dot{\phi}^2\ll
V\left(\kappa+2\beta\right).
\end{equation}
By taking the time derivative of Eq.~\eqref{HHdot}, one obtains
\begin{equation}\label{hdddott}
\ddot{H}=\dfrac{\dot{\phi}\ddot{\phi}}{\alpha}\left(\kappa+\beta\right).
\end{equation}
Then, substituting Eqs.~\eqref{hdddott} and \eqref{HHdot} into
Eq.~\eqref{2sr} yields the second extended slow-roll parameter as
\begin{equation}\label{etascalarrrr}
\eta\approx\dfrac{-\dfrac{\dot{\phi}\ddot{\phi}}{\alpha}\left(\kappa+\beta
\right)}{\dfrac{H}{\alpha}\dot{\phi}^2\left(\kappa+\beta\right)}=-\dfrac{|\ddot{\phi}|}{H|\dot{\phi}|},
\end{equation}
where the condition ${\rm \epsilon}\ll 1$ has been used. The
result indicates that $\eta$ does~not take any modification in
terms of derivatives of the scalar field. Also, by applying the
condition $|\eta|\ll 1$ on Eq.~\eqref{kgp}, one obtains the
time-derivative of the scalar field as
\begin{equation}\label{kgsrr}
\dot{\phi}\approx-\dfrac{\left(\kappa+2\beta\right)V^{\prime}}{3H\left(\kappa+\beta\right)}.
\end{equation}
Furthermore, by the modified slow-roll condition,
relation~\eqref{exsr}, Eq.~\eqref{HH2} leads to
\begin{equation}\label{exHsr}
H^2\approx-\dfrac{\kappa+2\beta}{3\alpha}V.
\end{equation}
Under these situations, by again applying the slow-roll
condition~\eqref{exsr}, into relation~\eqref{epssss} while using
Eqs.~\eqref{kgsrr} and \eqref{exHsr}, the first extended slow-roll
parameter can be written in terms of the scalar field potential as
\begin{equation}\label{epsscalar}
{\rm \epsilon}\approx\dfrac{3\left(\kappa+\beta\right)\dot{\phi}^2}{2\left(\kappa+2\beta
\right)V}\approx\dfrac{\left(\kappa+2\beta\right)V^{\prime 2}}{6H^2\left(\kappa+\beta
\right)V}\approx\dfrac{-\alpha}{2\left(\kappa+\beta\right)}\left(\dfrac{V^{\prime}}{V}\right)^2.
\end{equation}
Also, by taking the time derivative of Eq.~\eqref{kgsrr} and
substituting the result into relation~\eqref{etascalarrrr} while
using the condition ${\rm \epsilon}\ll 1$ and Eqs.~\eqref{kgsrr}
and \eqref{exHsr}, we obtain $\eta$ in terms of the scalar field
potential as
\begin{equation}\label{etascalar}
\eta\approx\dfrac{-\alpha}{\left(\kappa+\beta\right)}\left(\dfrac{V^{\prime\prime}}{V}\right).
\end{equation}
Moreover, once again by using the modified slow-roll
approximation~\eqref{exsr}, the effective parameter of the
equation of state, relation~\eqref{weff}, is $w^{[\rm
eff]}\approx-1$. Furthermore, by using Eqs.~\eqref{kgsrr} and
\eqref{exHsr}, the e-folding number for this model is
\begin{equation}\label{nv}
N\equiv\int_{\phi}^{\phi_{\rm
end}}\dfrac{H}{\dot{\phi}}d\phi\approx\dfrac{\left(\kappa+\beta
\right)}{-\alpha}\int_{\phi_{\rm
end}}^{\phi}\dfrac{V}{V^{\prime}}d\phi.
\end{equation}

Comparing relations~\eqref{epsscalar}, \eqref{etascalar} and
\eqref{nv} with relations \eqref{epsv}, \eqref{etav} and
\eqref{nGR} shows that the slow-roll parameters in the linear
$f(Q,T)$ gravity in this model with respect to those extracted
from GR take a modification as
\begin{equation}\label{modified kapp}
1/\kappa\rightarrow-\alpha/\left(\kappa+\beta\right),
\end{equation}
where $-\alpha/(\kappa+\beta)$ must be positive. This condition
obviously includes ($\alpha >0$ with $\beta <-k$) and ($\alpha <0$
with $\beta >-k$). Such the constant scaling of the $\kappa$ might
give a better chance that the linear $f(Q,T)$ gravity describe the
observational data better than GR does. Indeed, for different
values of the parameters due to such scaling, there might occur
some slight differences that may yield better fittings.

In the following, we investigate the linear $f(Q, T)$
gravitational model for a few different types of scalar field
potentials, and then specify the inflationary observables and
check their compatibility with the observational data.

\subsection{Power-Law Potential}
Let us consider the most simple type of scalar field potential
known as the power-law potential that leads to the chaotic
inflation~\cite{chaotic, powerlaw} and has the form
\begin{equation}\label{powerpotential}
V(\phi)=\nu\phi^n,
\end{equation}
where $\nu$ and $n$ are constants. For this potential, from
relations~\eqref{epsscalar} and \eqref{etascalar}, the slow-roll
parameters are
\begin{align}
\label{epspn}&{\rm\epsilon}\approx\dfrac{-\alpha\, n^2}{2\left(\kappa+\beta\right)\phi^2},\\
\label{etan}&\eta\approx\dfrac{-\alpha\,
n\left(n-1\right)}{\left(\kappa+\beta\right)\phi^2}.
\end{align}
Since inflation ends when the first slow-roll parameter reaches
the unit, the scalar field at the end of inflation can be
approximated as
\begin{equation}\label{phif}
\phi_{\rm end}^2\approx\dfrac{-\alpha\,
n^2}{2\left(\kappa+\beta\right)}.
\end{equation}

It is instructive to rewrite the slow-roll parameters in terms of
the e-folding number. For this purpose, first by substituting
Eq.~\eqref{powerpotential} into relation~\eqref{nv}, it reads
\begin{equation}\label{nfi}
N\approx\dfrac{-\left(\kappa+\beta\right)}{2\alpha\,
n}\left(\phi^2-\phi_{\rm end}^2\right).
\end{equation}
Then, by substituting relation~\eqref{phif} into it, the scalar
field at any time during the inflationary phase is
\begin{equation}\label{phi}
\phi^2\approx\dfrac{-\alpha\, n\left(4N+
n\right)}{2\left(\kappa+\beta\right)}.
\end{equation}
Finally, by substituting this result into relations~\eqref{epspn}
and \eqref{etan}, we obtain
\begin{align}
\label{epsi}&{\rm\epsilon}\approx\dfrac{ n}{4N+ n},\\
\label{etai}&\eta\approx\dfrac{2\left(n-1\right)}{4N+ n}.
\end{align}
These relations indicate that, in the case of power-law potential,
the slow-roll parameters do~not depend on the free parameters
$\alpha$ and $\beta$, and accordingly have no modification
compared to the results obtained~\cite{Gamonal:2020itt} from GR.
This is expected, since this case is equivalent to pure GR with a
minimally coupled scalar field (see the Appendix) that scales the
multiplicative constant $\nu$ of the power-law potential.

\begin{figure*}[t!]
\centering
\subfigure[]{
\includegraphics[width=0.31\textwidth]{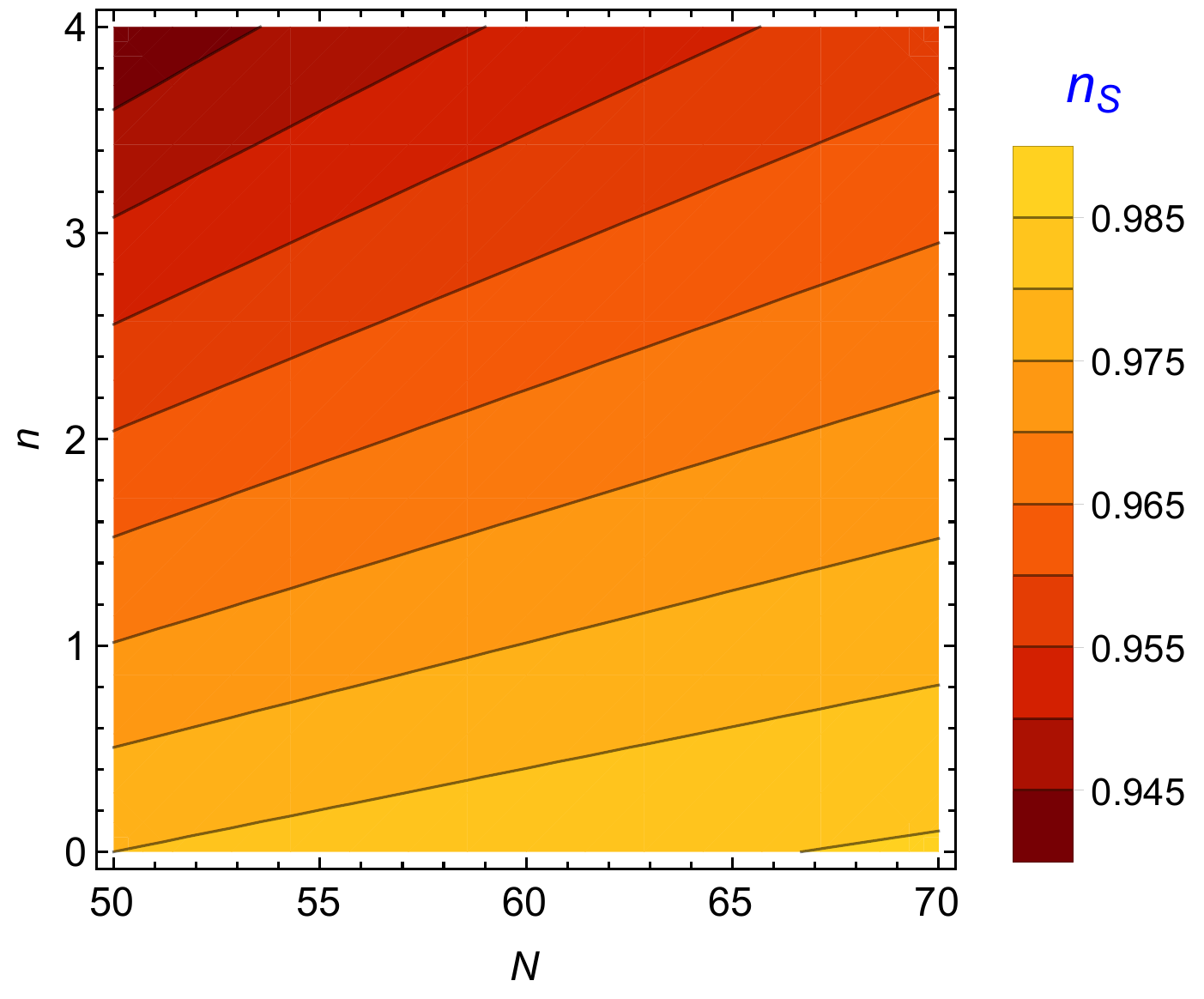}
\label{aa2}
}
\subfigure[]
{
\includegraphics[width=0.33\textwidth]{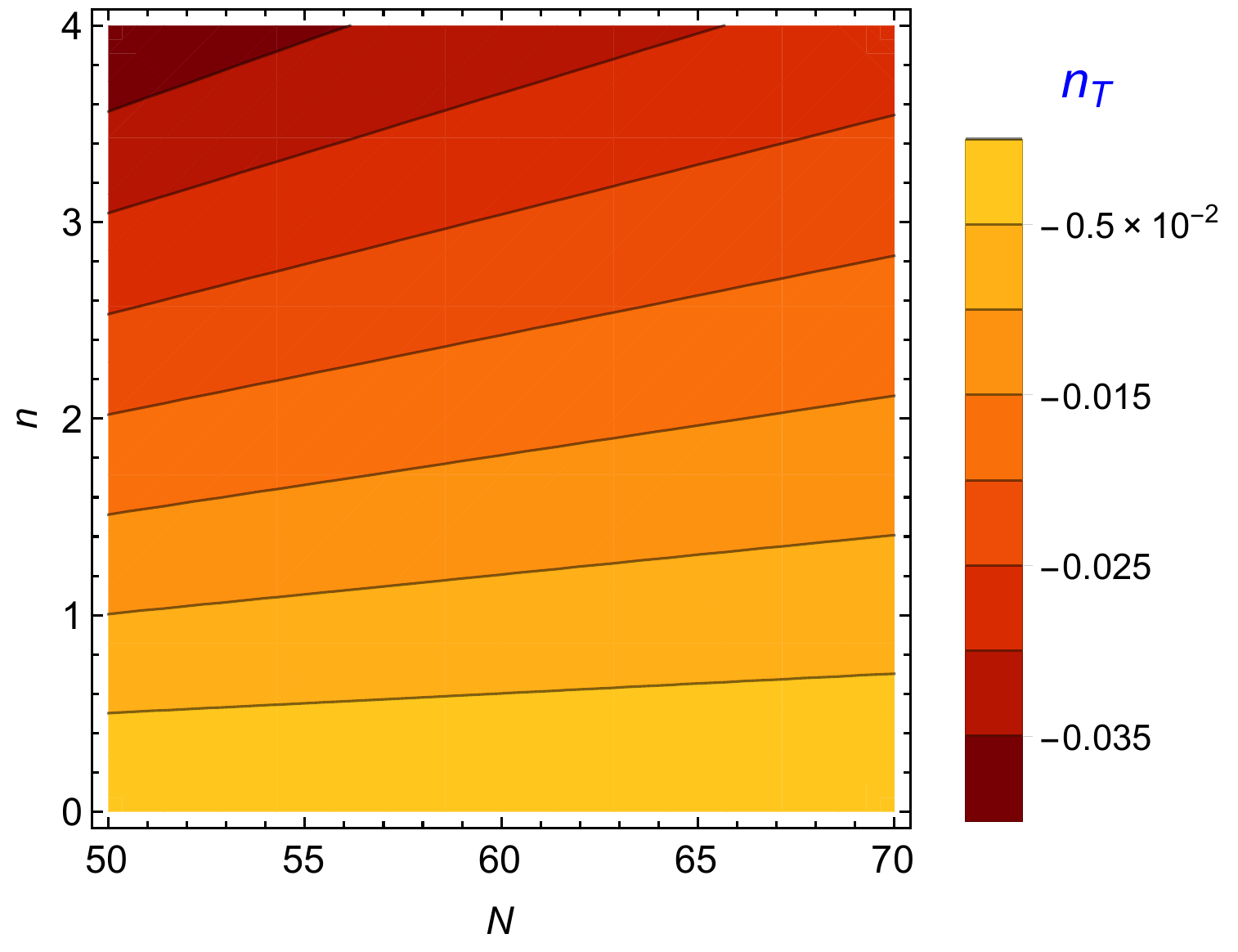}
\label{bb2}
}
\subfigure[]
{
\includegraphics[width=0.31\textwidth]{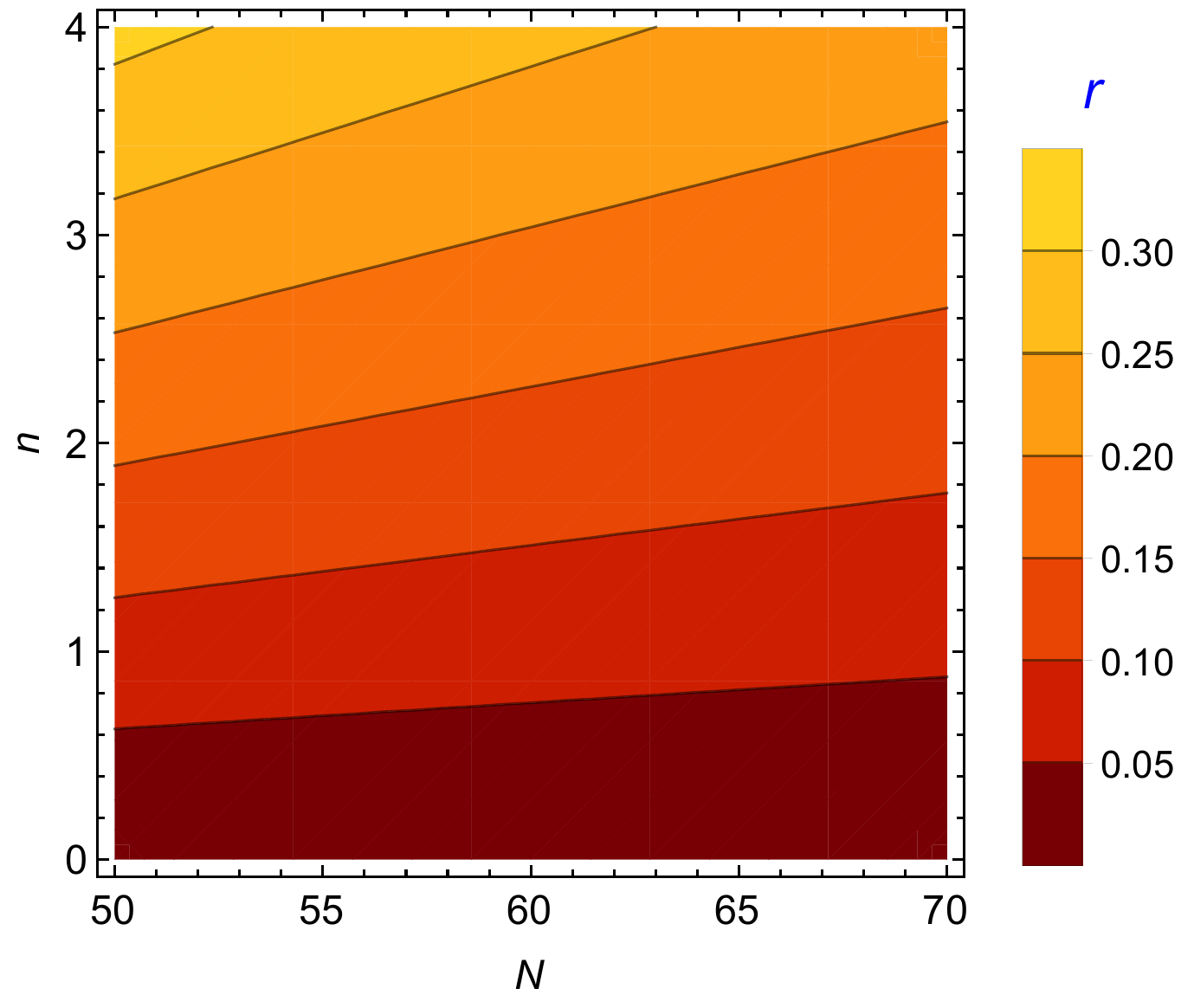}
\label{cc2} } \caption{[color online] Inflationary observables for
the power-law potential as functions of the power $n$ and the
e-folding number, wherein (a) shows the scalar spectral index,
$n_{\rm S}$, (b) indicates the tensor spectral index, $n_{\rm T}$,
and (c) demonstrates the tensor-to-scalar ratio, $r$.}
\label{fig2}
\end{figure*}

Also, using the inflationary observables defined in
relations~\eqref{ns}, \eqref{nt} and \eqref{r}, one obviously
obtains
\begin{align}
&n_{\rm S}\approx\dfrac{4\left(N-1\right)-n}{4N+n},\\
&n_T\approx\dfrac{-2n}{4N+ n},\\
&r\approx\dfrac{16 n}{4N+n}.
\end{align}
Hence, the inflationary observables do~not depend on the free
parameters of the model as well. The only key parameters in
calculating these observations are the power $n$ and the e-folding
number. In this regard, the numerical results for $n_{\rm S}$,
$n_{\rm T}$ and $r$ for different values of $n$ and $N$ have been
demonstrated in Fig.~\ref{fig2} in the range of $[0,4]$ and
$[50,70]$, respectively. Also, the consistent results for some
values of $n$ and $N$ have been presented in Table~\ref{table2}.
Obviously, these results are in good agreement with the latest
observational data obtained from the Planck satellite.

For the case of $n=2$, the power-law potential is related to the
free field also known as the Klein-Gordon potential. In this case,
also $\nu=m^{2}/2$, where $m$ is the mass of the scalar field. The
results indicate that the acceptable range for $(n_{\rm S},
n_{T})$ is $(0.96460,0.14159)$ with $N=56$. Also, for $n=4$, the
power-law potential corresponds to the Higgs potential. For this
case, $\nu=\lambda/4$, where $\lambda$ is the coupling constant,
and the appropriate range for $(n_{\rm S}, n_{T})$ is
$(0.95774,0.22535)$ with $N=70$.

\begin{table*}[t]
\caption{Inflationary observables in the power-law inflation for
some values of $n$ and $N$.} \centering
\begin{tabular}{ c | c | c | c | c }
\hline
\hline
$n$ & $N$ & $n_{\rm S}$ & $r$ & $n_{\rm T}$ \\ [0.5ex]
\hline
0.5 & 50  & 0.97506 & 0.03990 & -0.00498 \\
1.0 & 50  & 0.97014 & 0.07960 & -0.00995 \\
1.5 & 51  & 0.96593 & 0.11678 & -0.01459 \\
2.0 & 56  & 0.96460 & 0.14159 & -0.01769 \\
2.5 &  62 & 0.96407 & 0.15968 & -0.01996 \\
3.0 &  70 & 0.96466 & 0.16961 & -0.02120 \\
3.5 &  70 & 0.96119 & 0.19753 & -0.02469 \\
4.0 &  70 & 0.95774 & 0.22535 & -0.02816 \\
\hline
\hline
\end{tabular}
\label{table2}
\end{table*}
\subsection{Hyperbolic Potential}
In Ref.~\cite{rubanoandbarrow}, an exact hyperbolic form of the
scalar field potential has been specified at the spatially flat
Friedmann Universe containing a scalar field with equation of
state $p^{[\phi]}=\left(\gamma^{[\phi]}-1\right)\rho^{[\phi]}$ and
a perfect fluid with equation of state $p^{[\rm
m]}=\left(\gamma-1\right)\rho^{[\rm m]}$ (where
$\gamma^{[\phi]}=w^{[\phi]}+1$ and $\gamma=w+1$), in terms of
$H_0$, $\Omega^{[\rm m]}_0$, $w$ and $w^{[\phi]}$. Also, dark
energy has been discussed with cosmological acceleration from the
scalar field with such a potential at the late-time Universe.
Moreover, in Ref.~\cite{hyperbolic}, the hyperbolic scalar field
potential has been studied in the concept of inflation in GR,
which leads to consistent inflationary observables with the Planck
$2015$ data for some ranges of $h$ and $b$ parameters, and derives
$r\simeq0.07$ when $26 {\rm M_{Pl}} \leq h \leq 39 {\rm M_{Pl}}$
and $1.02 \leq b \leq 1.1$. In this work, we intend to investigate
the hyperbolic potential within the context of $f(Q, T)$ gravity.

The hyperbolic potential can be described as
\begin{equation}\label{hypp}
V(\phi)=A\,\sinh^b\left(\dfrac{\phi}{h}\right),
\end{equation}
where $h$ is scaled as the Planck mass, and the constants $A$ and
$b$ have been defined as~\cite{rubanoandbarrow}
\begin{align}
\label{A}&A\equiv
3H_0^2\left(1-\Omega^{[m]}_0\right)\left(1-\dfrac{\gamma^{[\phi]}}{2}
\right)\left(\dfrac{1-\Omega^{[m]}_0}{\Omega^{[m]}_0}\right)^{-b/2},\\
\label{b}&b\equiv
-\dfrac{2\gamma^{[\phi]}}{\gamma-\gamma^{[\phi]}}=\dfrac{2\left(1+w^{[\phi]}\right)}{1+w^{[\phi]}-\gamma}.
\end{align}
In these relations, $H_0$ and $\Omega^{[m]}_0$ are the value of
the present-day Hubble parameter and the present-day matter
density parameter, respectively. Obviously, for the
radiation-dominated era, the constant $b$ is
\begin{equation}
b=\dfrac{6\left(1+w^{[\phi]}\right)}{3w^{[\phi]}-1}.
\end{equation}
Substituting potential~\eqref{hypp} into
relations~\eqref{epsscalar} and \eqref{etascalar} gives the
slow-roll parameters for this case to be
\begin{equation}
\label{epshyp}\epsilon\approx\dfrac{-\alpha\,
b^2}{2\left(\kappa+\beta\right)h^2}\coth^2\left(\dfrac{\phi}{h}\right),
\end{equation}
\begin{equation}
\label{etahyp}\eta\approx\dfrac{-\alpha\,
b}{\left(\kappa+\beta\right)h^2}\dfrac{\left[b\cosh^2
\left(\dfrac{\phi}{h}\right)-1\right]}{\left[\cosh^2\left(\dfrac{\phi}{h}\right)-1\right]}.
\end{equation}
Also, by using the end of inflation condition,
${\rm\epsilon}(\phi_{{\rm end}})=1$, we obtain
\begin{equation}\label{phiendhyp}
\phi_{{\rm end}}\approx h\times\,{\rm
arccoth}\left[\sqrt{\dfrac{-2\left(\kappa+\beta\right)h^2}{\alpha\,
b^2}}\right],
\end{equation}
where again $-\alpha/(\kappa+\beta)$ must be positive. Moreover,
by inserting potential~\eqref{hypp} into relation~\eqref{nv}, the
e-folding number reads
\begin{equation}\label{nhypefold}
N\approx\dfrac{-h^2\left(\kappa+\beta\right)}{\alpha\,
b}\ln\left[\dfrac{\cosh\left(\dfrac{\phi}{h}
\right)}{\cosh\left(\dfrac{\phi_{\rm end}}{h}\right)}\right].
\end{equation}
Consequently, by substituting relation~\eqref{phiendhyp} into
relation~\eqref{nhypefold}, we obtain the scalar field as
\begin{equation}\label{phihyp}
\phi\approx h\times\,{\rm
arccosh}\left[\dfrac{\exp\left(\dfrac{-\alpha\, b
N}{\left(\kappa+\beta
\right)h^2}\right)\sqrt{\dfrac{-2\left(\kappa+\beta\right)h^2}{\alpha\,
b^2}}}{\sqrt{\dfrac{-2 \left(\kappa+\beta\right)h^2}{\alpha\,
b^2}-1}}\right],
\end{equation}
where $\left[-2 \left(\kappa+\beta\right)h^2/(\alpha\,
b^2)-1\right]$ must be positive. This condition makes more
restriction on $\alpha$ and $\beta$ than the above one (under
relation~\eqref{phiendhyp}), e.g. it includes [$\alpha >0$ with
$\beta <-\kappa-\alpha\,b^2/(2h^2)$] and [$\alpha <0$ with $\beta
>-\kappa-\alpha\,b^2/(2h^2)$]. Under these conditions, the slow-roll parameters read
\begin{equation}
\epsilon\approx\dfrac{-\alpha\, b^2 }{{\rm
D}}\exp\left[\dfrac{-2\alpha\,
bN}{\left(\kappa+\beta\right)h^2}\right],
\end{equation}
\begin{equation}
\eta\approx\dfrac{-\alpha b \left\lbrace
2h^2\left(\kappa+\beta\right)\!\left[b\exp \left(\dfrac{-2\alpha b
N}{\left(\kappa +\beta\right)h^2}\right)\!-1\right]\!-\alpha b^2
\right\rbrace }{h^2 {\rm D}\left(\kappa +\beta\right)},
\end{equation}
where
\begin{equation}
{\rm D}\equiv 2h^2\left(\kappa+\beta\right)\left[\exp\left(\dfrac{-2\alpha bN}
{\left(\kappa+\beta\right)h^2}\right)-1\right]-\alpha b^2.
\end{equation}
As a result, the inflationary observables, in this case, obviously
are
\begin{align}
&n_{\rm S}\approx 1+\dfrac{6\alpha b^2}{{\rm D}}\exp\left(\dfrac{-2\alpha bN}{\left(\kappa+\beta\right)h^2}\right)\nonumber\\
&-\dfrac{2\alpha b \left\lbrace
2h^2\left(\kappa+\beta\right)\left[b\exp \left(\dfrac{-2\alpha b
N}{\left(\kappa +\beta\right)h^2}\right)-1\right]-\alpha b^2
\right\rbrace }{h^2 {\rm D}\left(\kappa+\beta\right)},
\end{align}
\begin{equation}
n_T\approx\dfrac{2\,\alpha\, b^2}{{\rm
D}}\exp\left[\dfrac{-2\,\alpha\,
b\,N}{\left(\kappa+\beta\right)h^2}\right],
\end{equation}
\begin{equation}
r\approx\dfrac{-16\,\alpha\, b^2 }{{\rm
D}}\exp\left[\dfrac{-2\,\alpha\,
b\,N}{\left(\kappa+\beta\right)h^2}\right].
\end{equation}
As is clear from these relations, within the framework of the
linear functional form of $f(Q, T)$ gravity while considering the
hyperbolic scalar field potential, the inflationary observables
depend on parameters $b$, $N$ and on a single combination of the
other parameters, i.e. $-h^2(\kappa+\beta)/\alpha$, with respect
to GR. The scaling change
\begin{equation}\label{hScaling}
h^2 \kappa \rightarrow \dfrac{h^2(\kappa+\beta)}{-\alpha},
\end{equation}
stems from relation~\eqref{modified kapp} that leads to the
relations resulting from GR with a minimally coupled scalar field
(see the Appendix). Such a modification can cause even slight
differences compared to the results of the GR case. For instance,
to obtain consistent results with the Planck data in GR, the
allowed range of $h$ is $h\geq 11.7 {\rm
M_{Pl}}$~\cite{hyperbolic}. In this case of linear $f(Q, T)$
gravity, relation\eqref{hScaling} can impose different
restrictions on the corresponding $h$ value, i.e. $\tilde{h}\geq
\sqrt{(\kappa+\beta)/(-\alpha\kappa)}\, 11.7 {\rm M_{Pl}}$, where
we have set $h^2\kappa \rightarrow \tilde{h}^2\kappa$.

\begin{figure*}[t!]
\centering
\subfigure[]{
\includegraphics[width=0.31\textwidth]{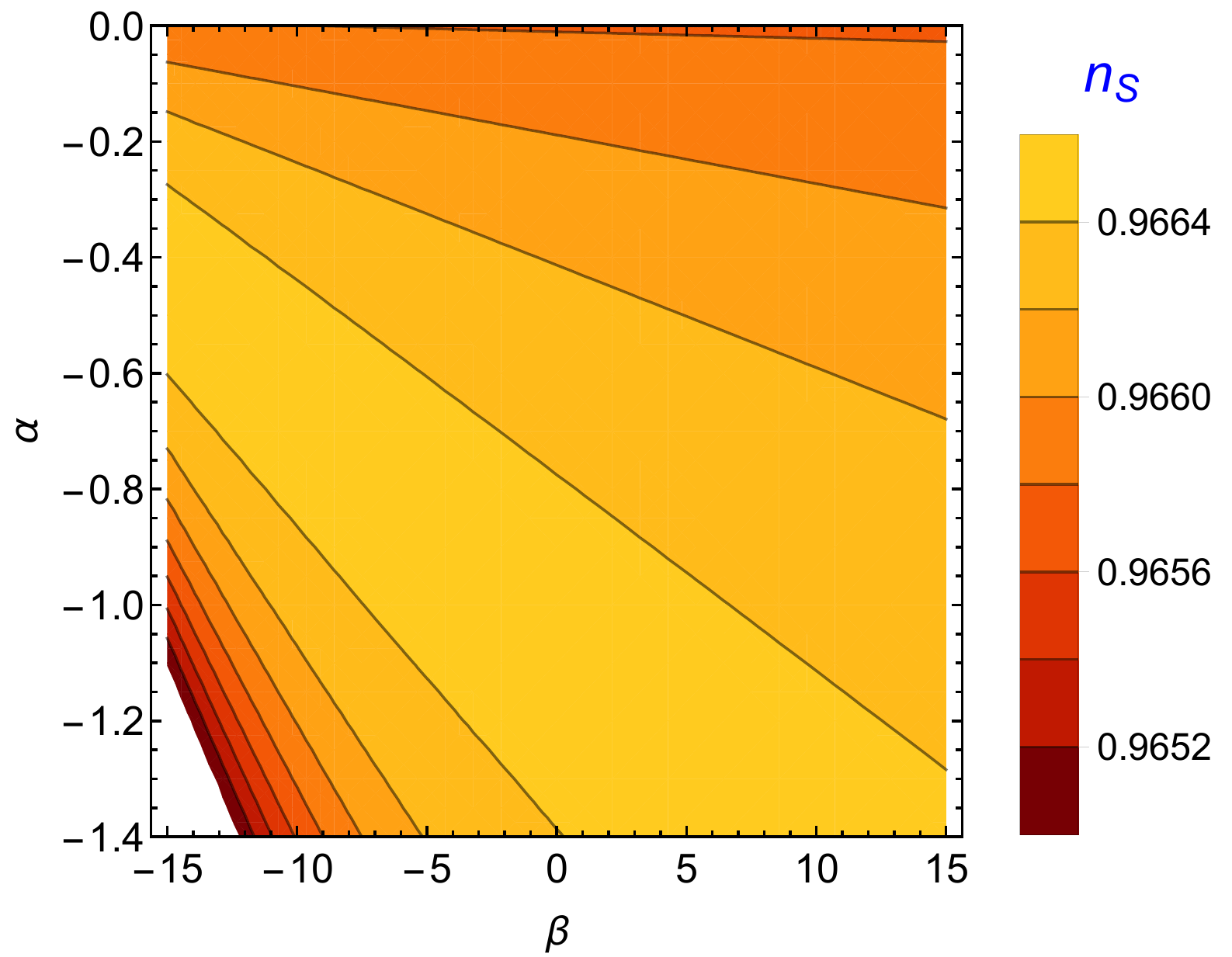}
\label{aa3}
}
\subfigure[]
{
\includegraphics[width=0.32\textwidth]{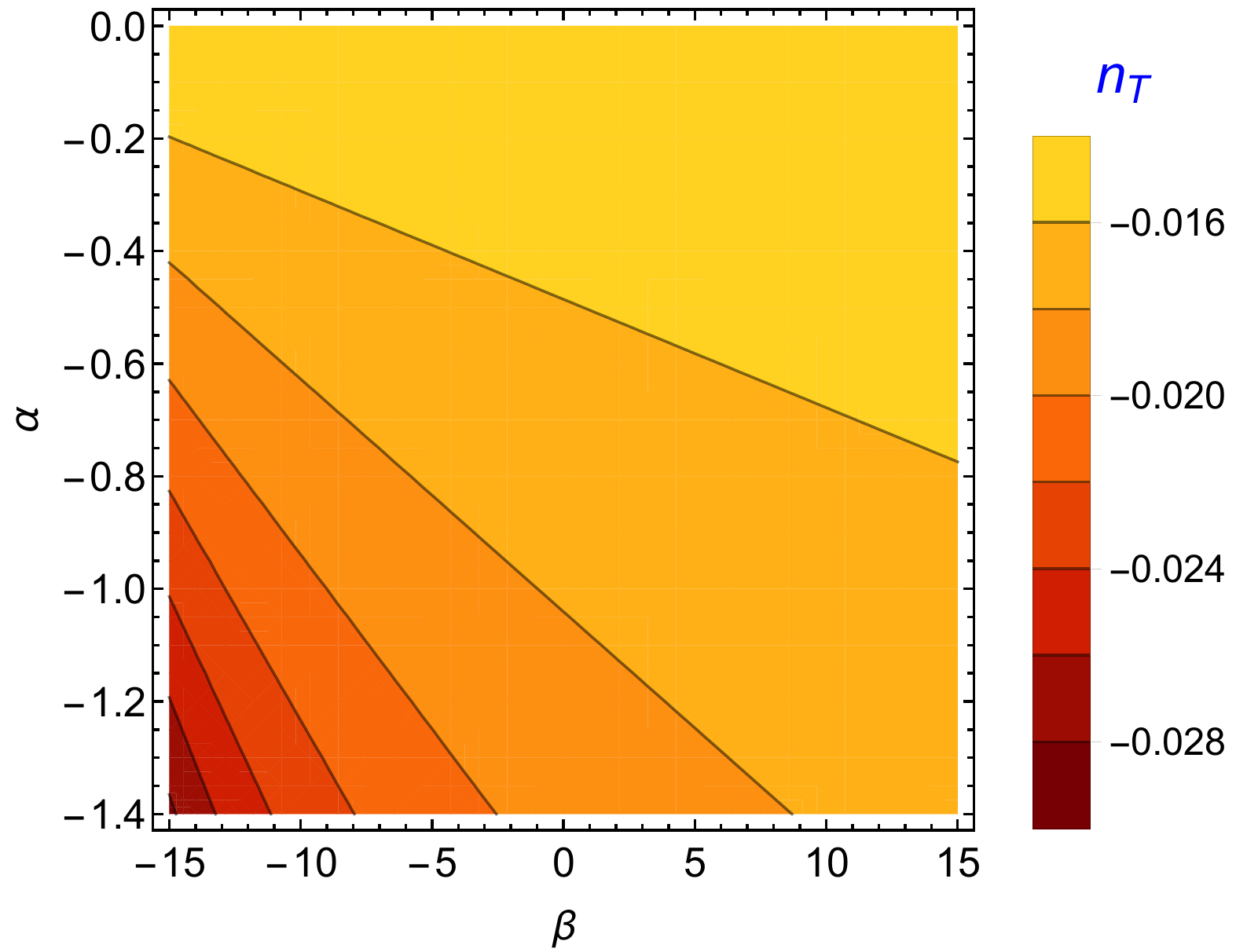}
\label{bb3}
}
\subfigure[]
{
\includegraphics[width=0.31\textwidth]{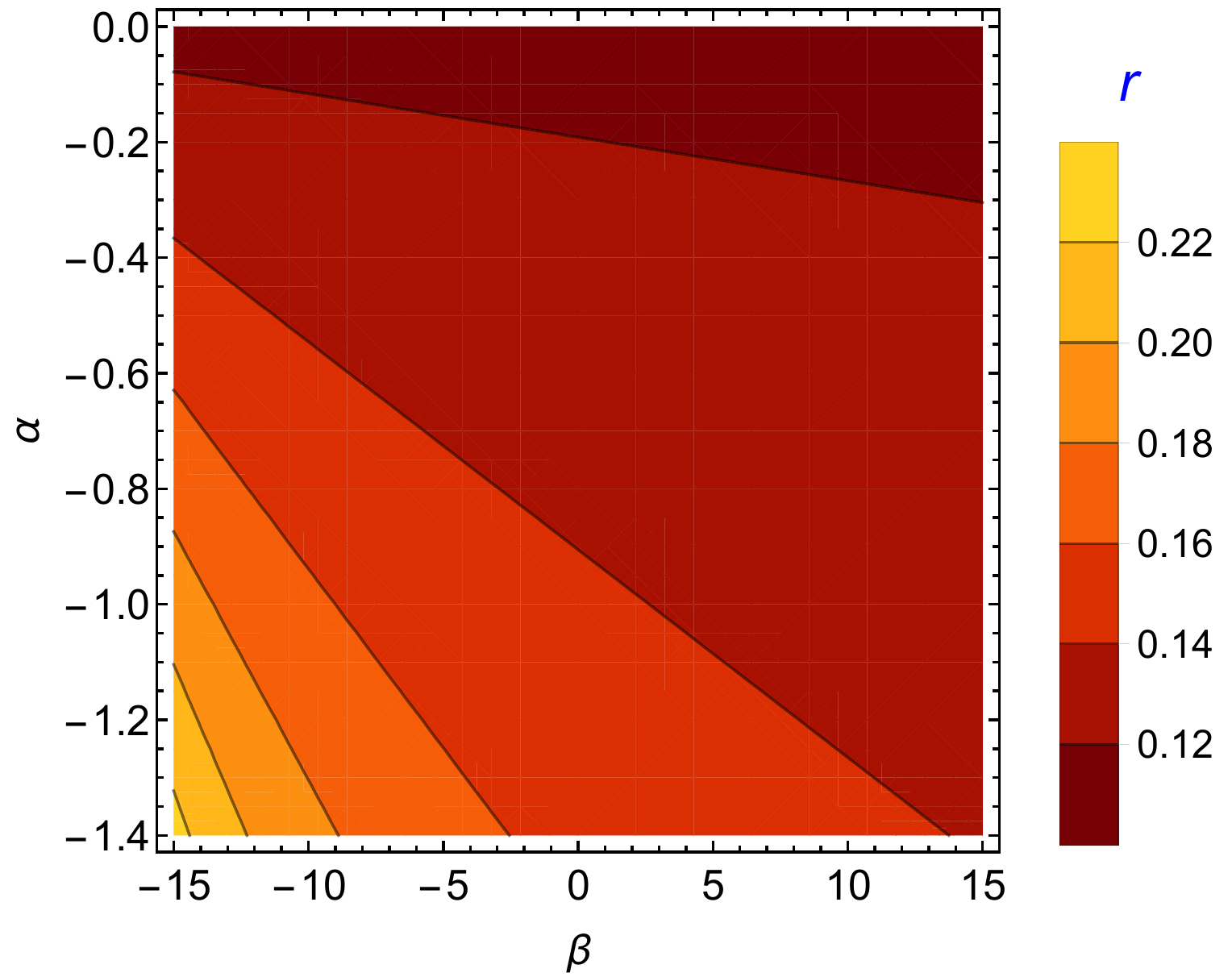}
\label{cc3}
}
\subfigure[]{
\includegraphics[width=0.31\textwidth]{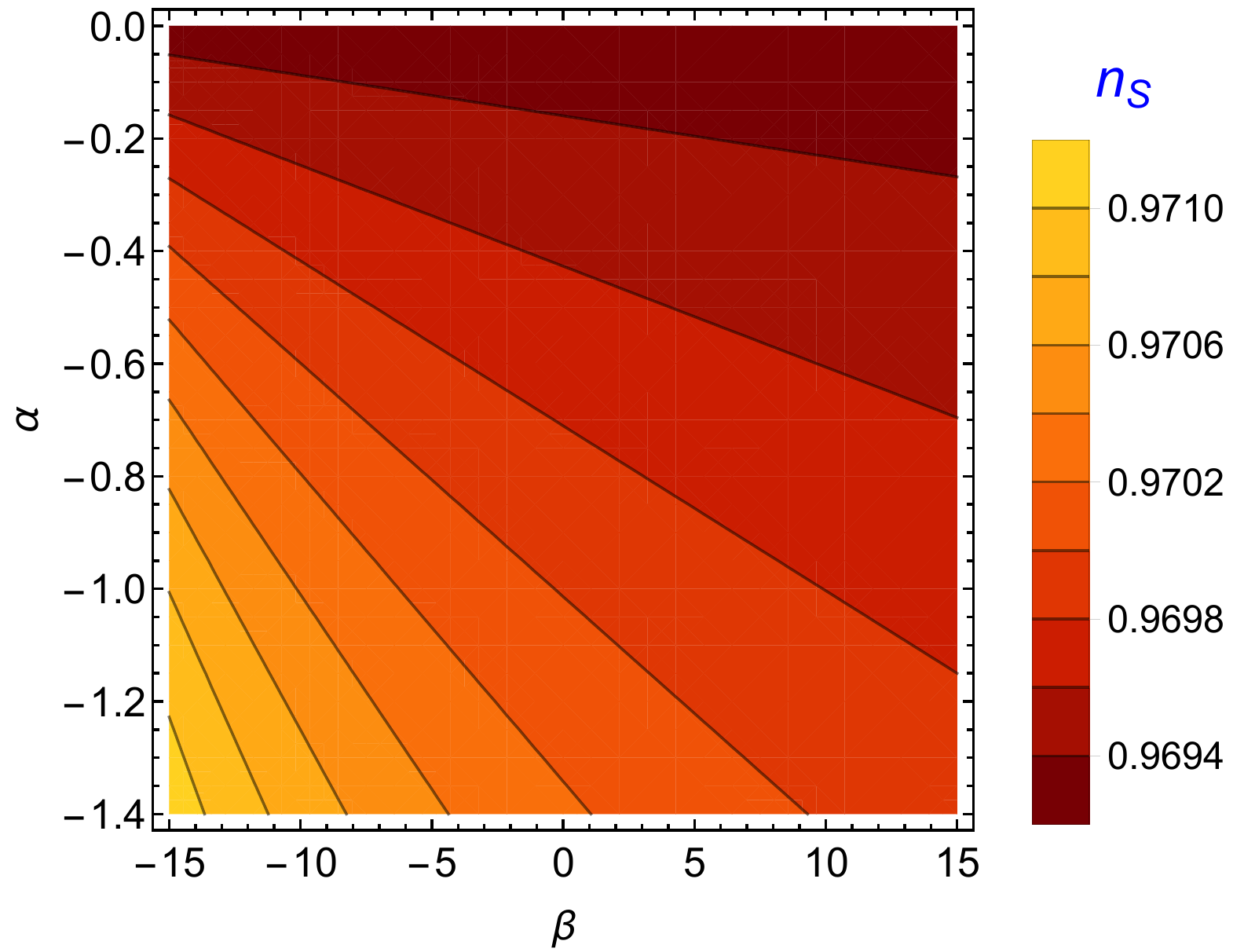}
\label{dd3}
}
\subfigure[]
{
\includegraphics[width=0.32\textwidth]{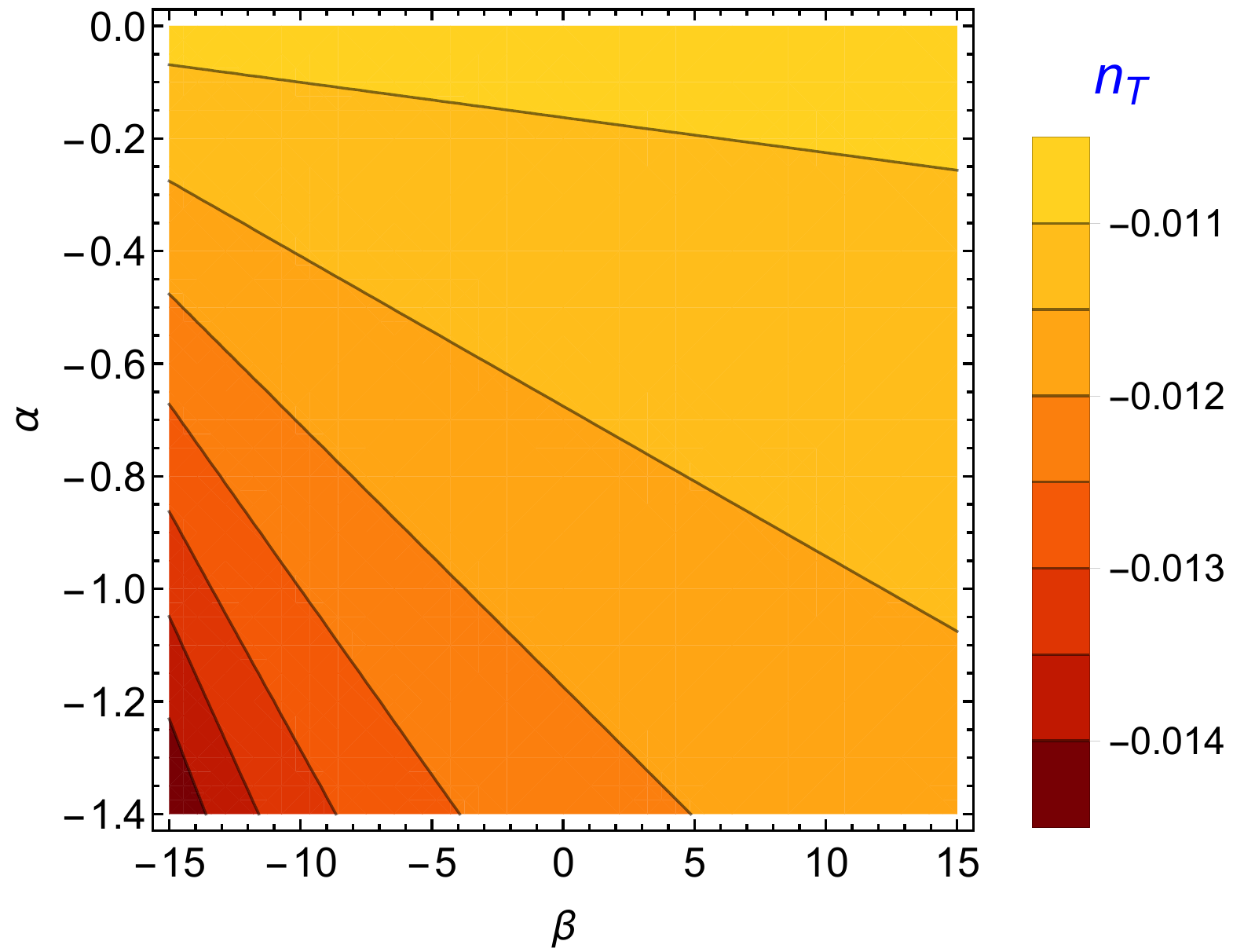}
\label{ee3}
}
\subfigure[]
{
\includegraphics[width=0.31\textwidth]{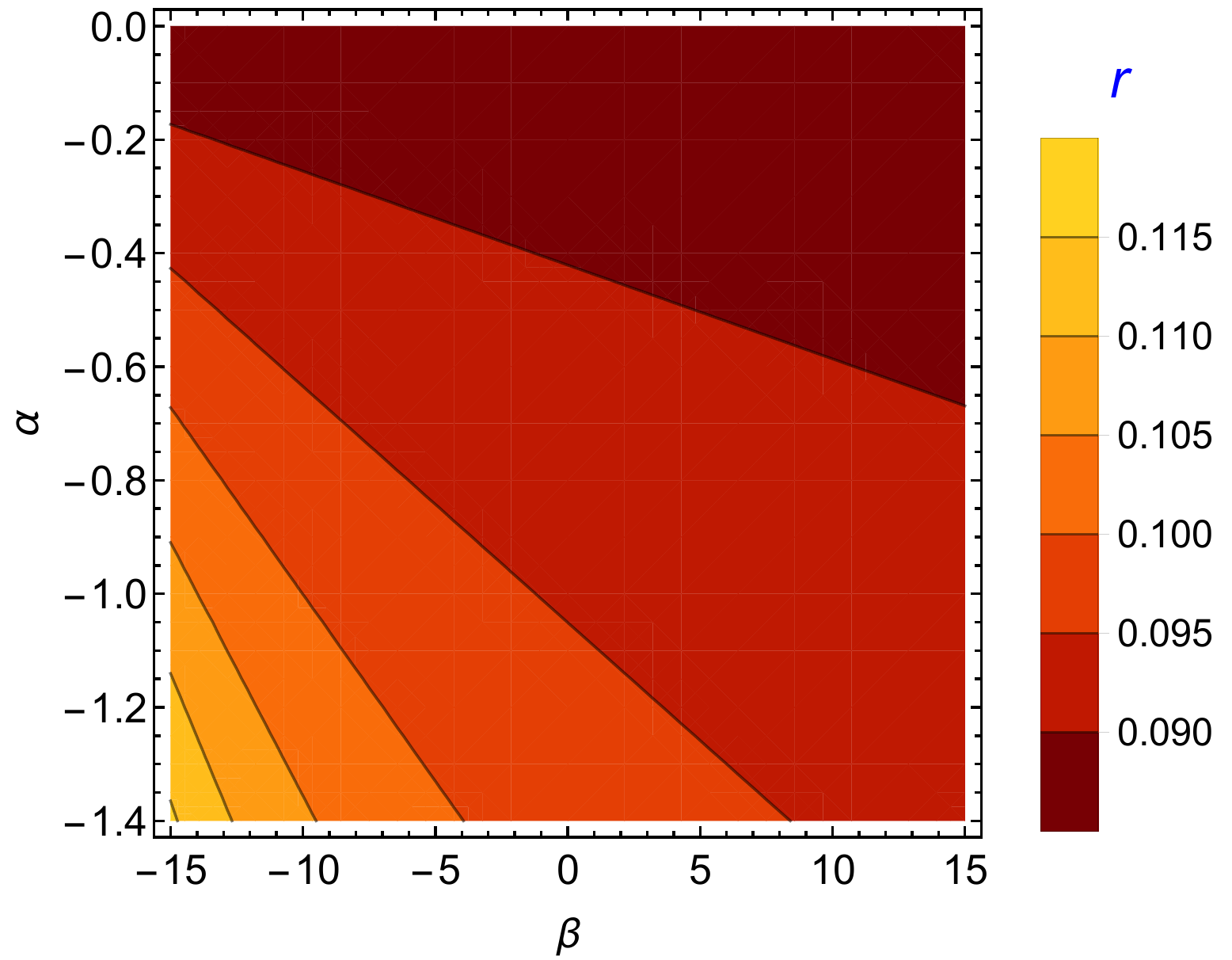}
\label{ff3} } \caption{[color online] Inflationary observables for
the case of hyperbolic potential as functions of parameters
$\alpha < 0$ and $\beta$. Top figures have been calibrated with
$h=18\,M_{\rm Pl}$, $b=1.448$ and $N=50$, and bottom figures with
$h=25\,M_{\rm Pl}$, $b=1.090$ and $N=50$.} \label{fig3}
\end{figure*}

\begin{figure*}[t!]
\centering
\subfigure[]{
\includegraphics[width=0.31\textwidth]{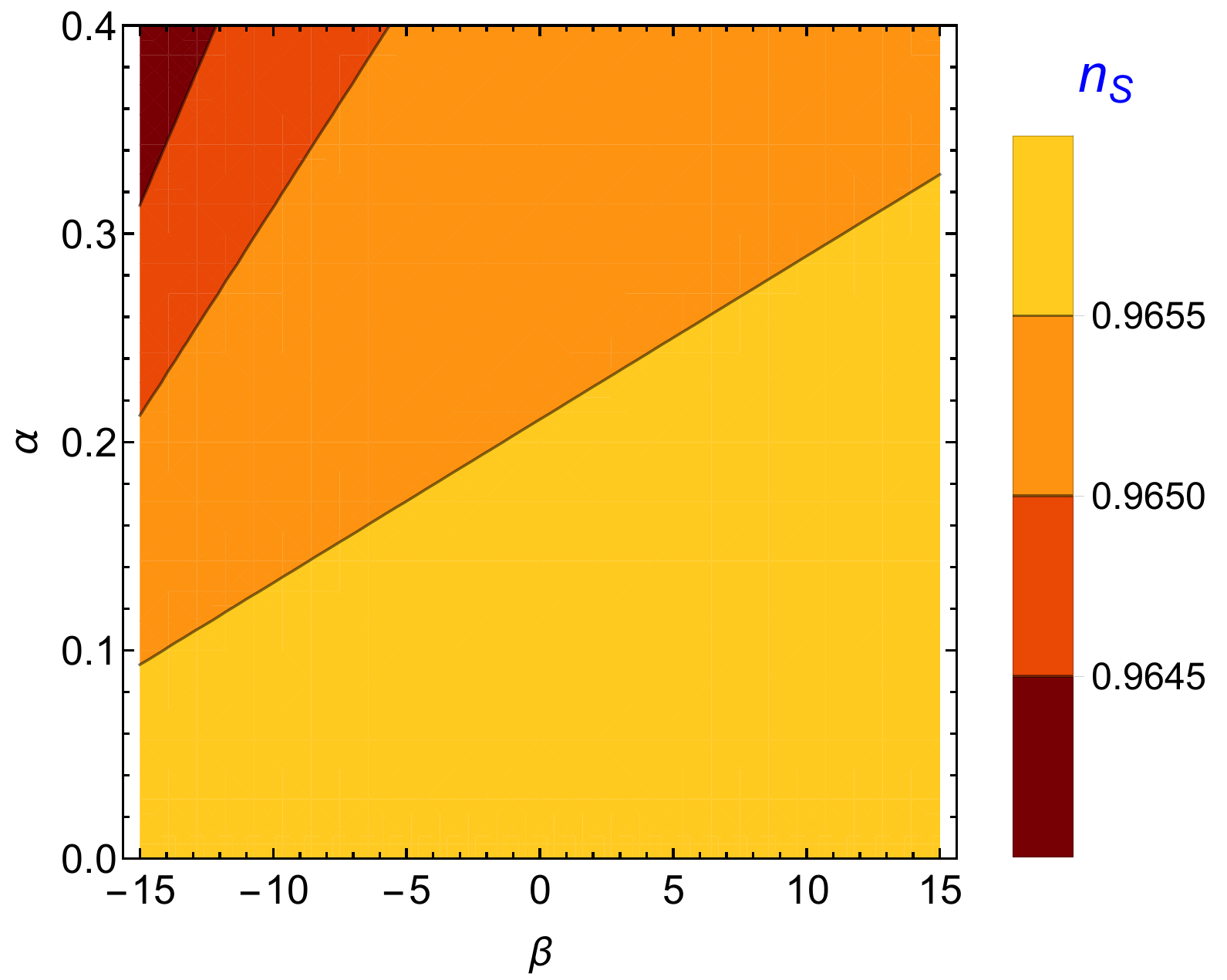}
\label{aa4}
}
\subfigure[]
{
\includegraphics[width=0.315\textwidth]{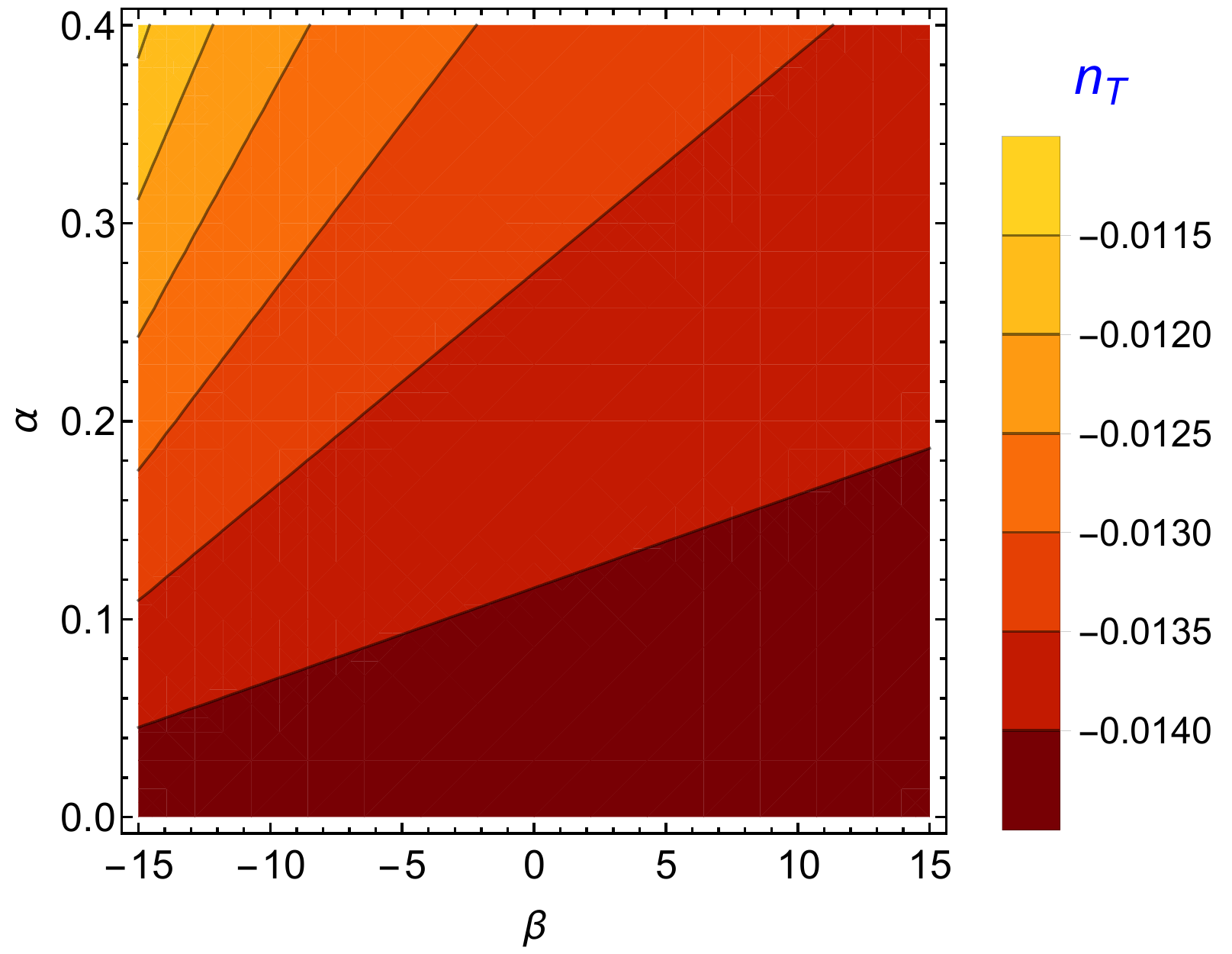}
\label{bb4}
}
\subfigure[]
{
\includegraphics[width=0.31\textwidth]{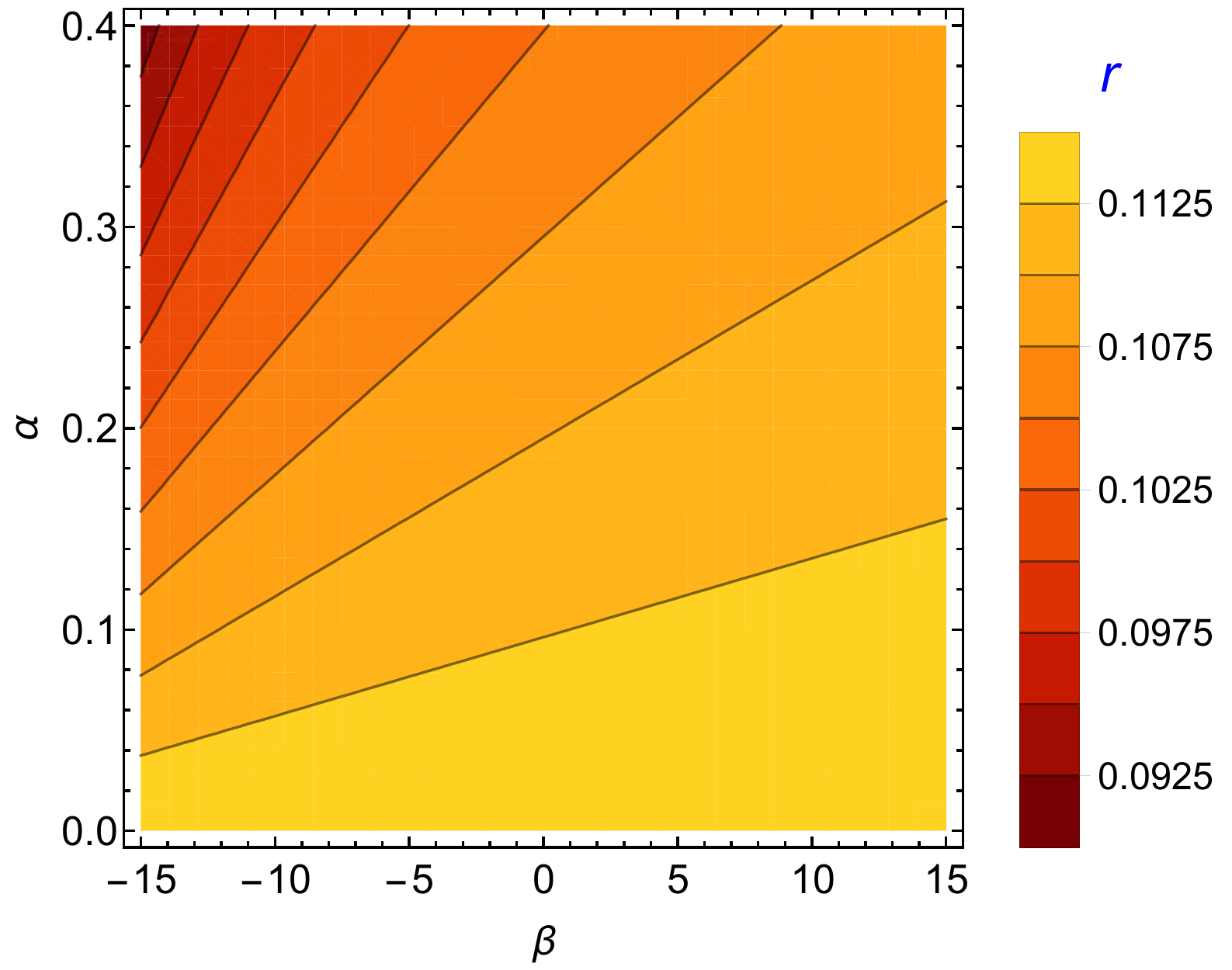}
\label{cc4}
}
\subfigure[]{
\includegraphics[width=0.30\textwidth]{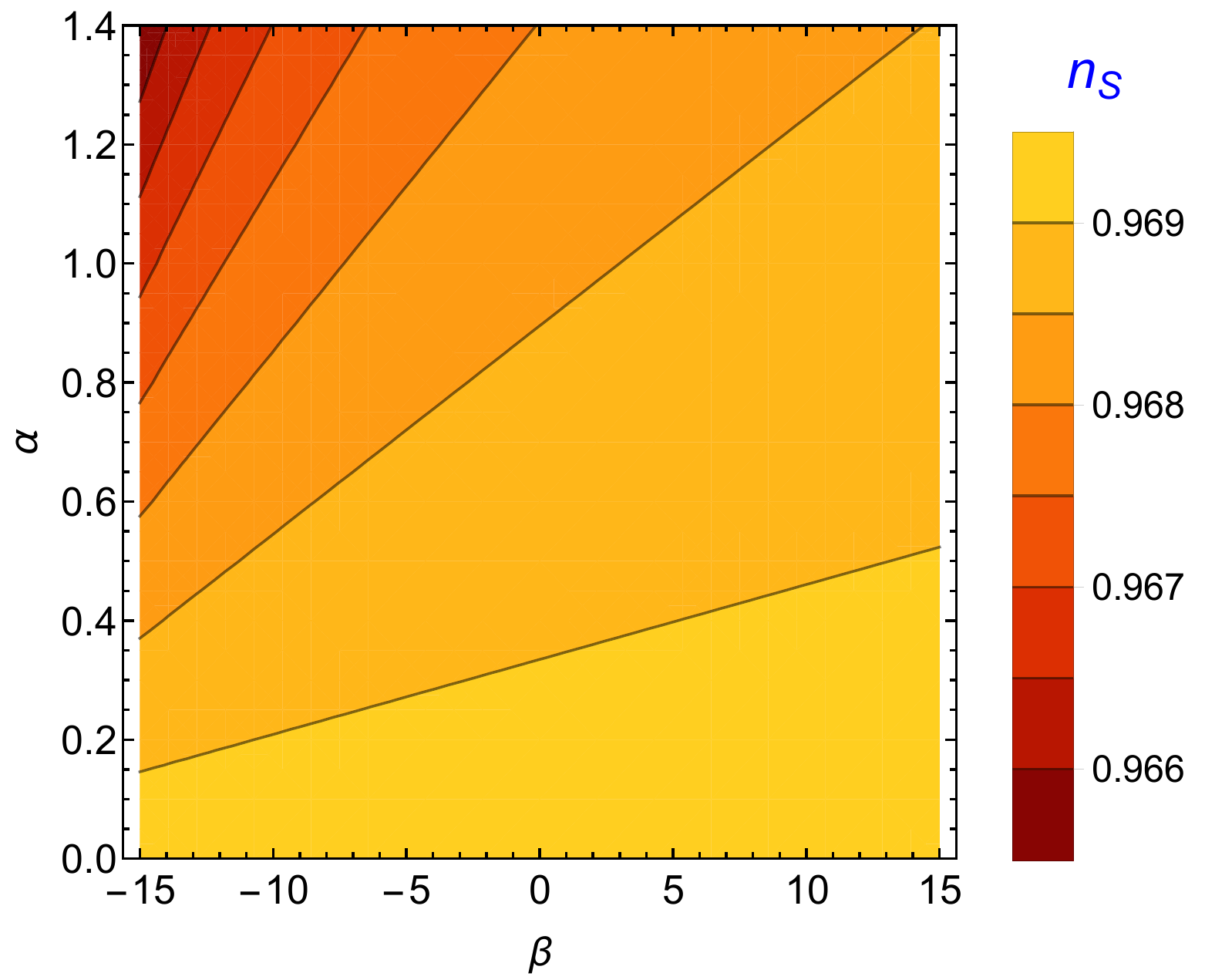}
\label{dd4}
}
\subfigure[]
{
\includegraphics[width=0.33\textwidth]{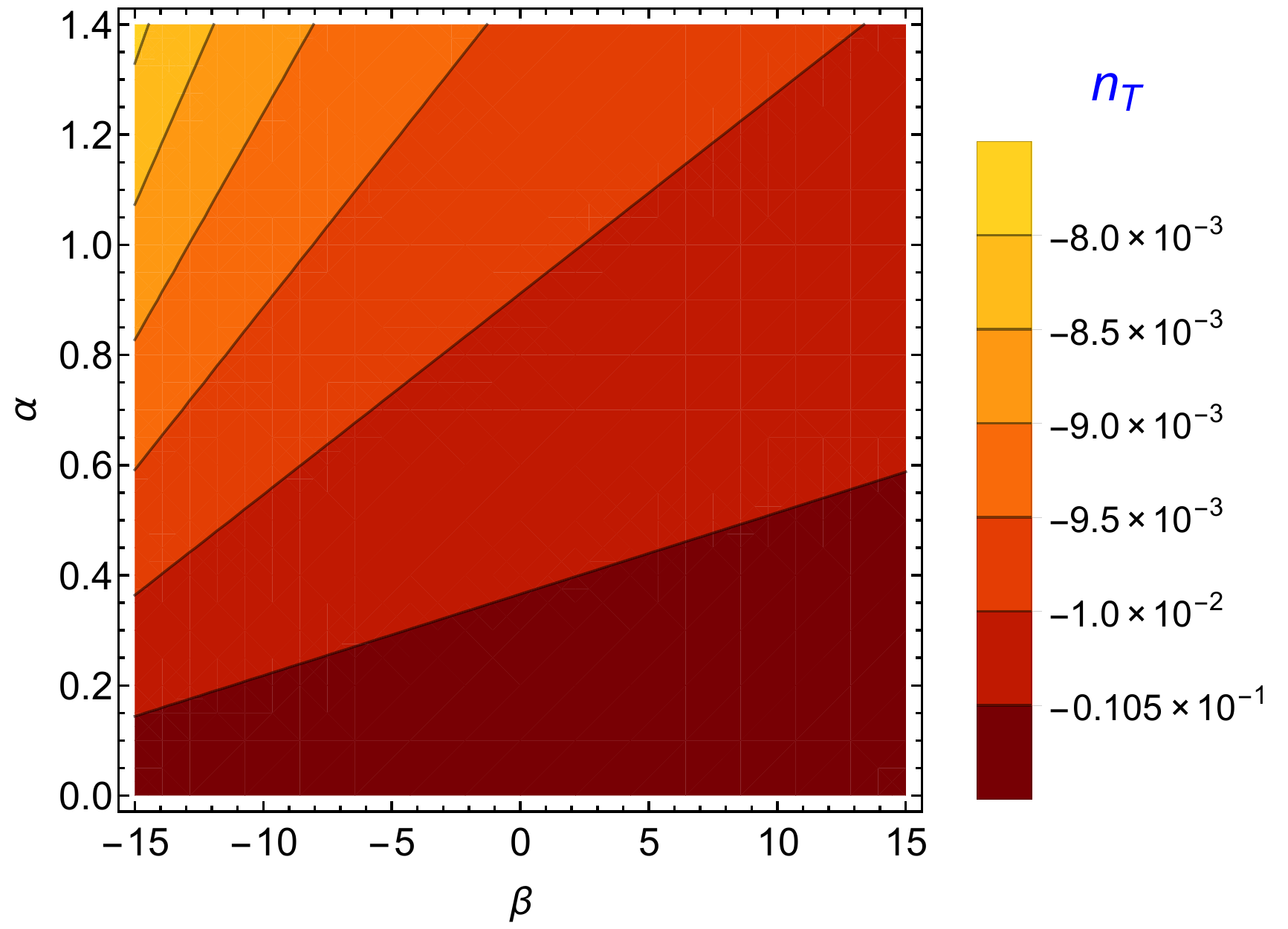}
\label{ee4}
}
\subfigure[]
{
\includegraphics[width=0.31\textwidth]{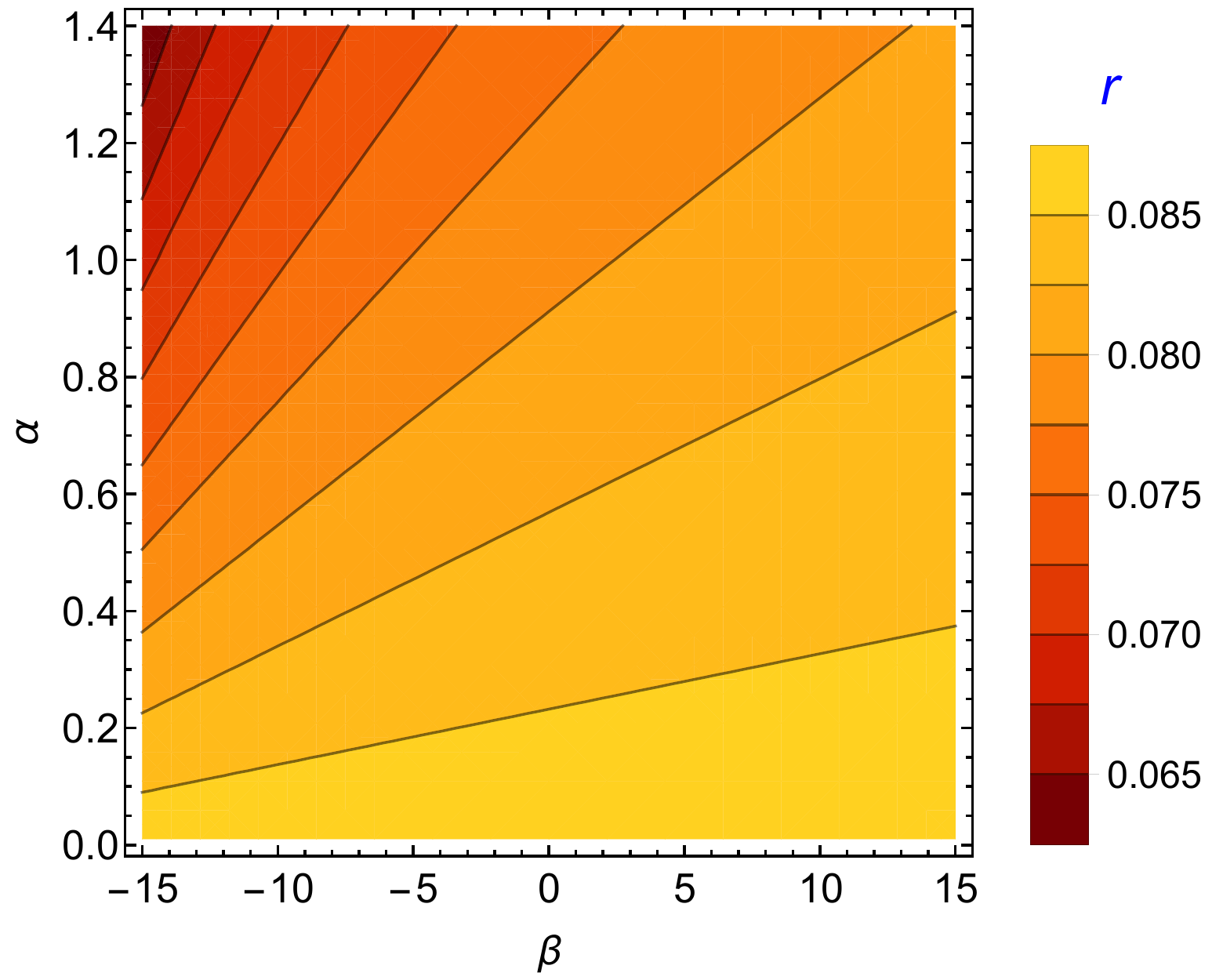}
\label{ff4} } \caption{[color online] Inflationary observables for
the case of hyperbolic potential as functions of parameters
$\alpha > 0$ and $\beta$. Top figures have been calibrated with
$h=18\,M_{\rm Pl}$, $b=1.448$ and $N=50$, and bottom figures with
$h=25\,M_{\rm Pl}$, $b=1.090$ and $N=50$.} \label{fig4}
\end{figure*}

Figs.~\ref{fig3} and \ref{fig4} show the corresponding
inflationary observables of the hyperbolic potential with $N=50$,
while considering ($b=1.448$ and $h=18M_{\rm Pl}$) and ($b=1.090$
and $h=25M_{\rm Pl}$) with different values of $\beta$, and
negative and positive values of $\alpha$, respectively. Also, some
acceptable values of the inflationary observables have been
presented in Table~\ref{table3} for different values of $h$, $b$,
$\alpha$, and $\beta$. The results confirm that the inflationary
observables in the $f(Q, T)$ gravity while considering the
hyperbolic potential are consistent with the latest Planck $2018$
observational data. However, for positive values of $\alpha$,
while relaxing the condition $-\alpha/\(\kappa+\beta\)>0$, we
obtain more consistent results with the joint Planck, BK15 and BAO
\eqref{planckdata2}. Moreover, it makes $r$ more restrictive than
the hyperbolic inflation in GR. Hence, the hyperbolic potential
almost leads to a viable $f(Q,T)$ gravity model in the early stage
of the Universe.

\begin{table*}[ht]
\caption{Inflationary observables in the hyperbolic inflation for
some values of $h$, $b$, $\alpha$ and $\beta$. Note that, for some
of the given values, the condition
$-\alpha/(\kappa+\beta)>0$ is relaxed in order to consider
further matter configurations.}
 \centering
\begin{tabular}{c | c | c | c | c | c | c}
\hline \hline $h$ (in $M_{\rm Pl}$) & $b$ & $\alpha$ & $\beta$ &
$n_{\rm S}$ & $r$ & $n_{\rm T}$ \\ [0.5ex] \hline
14 & 1.480 & 0.63 & 8.74 & 0.96418 & 0.09786 & -0.01223\\
14 & 1.480 & -0.97 & -7.94 & 0.96499 & 0.19206 & -0.02400\\
18 & 1.448 & 0.28 & -12.81 & 0.96491 & 0.10076 & -0.01259\\
18 & 1.448 & -1.31 & -13.22 & 0.96494 & 0.20090 & -0.02511\\
18 & 1.448 & -1.49 & -11.94 & 0.96479 & 0.20362 & -0.02511\\
25 & 1.090 & 1.21 & 3.19  & 0.96832  & 0.07884  & -0.00985\\
25 & 1.090 & 2.51 & -10.02  & 0.96460  & 0.05879  & -0.00734\\
25 & 1.090 & -0.15 &  2.51  & 0.96938  & 0.08780  & -0.01097\\
25 & 1.090 & -0.53 & -7.25 & 0.96983 & 0.09253 & -0.01156\\
30 & 1.009 & 3.68 & -11.35 & 0.96492 & 0.05371 & -0.00671\\
30 & 1.009 & 4.54 & -7.95 & 0.96498 & 0.05396 & -0.00674\\
30 & 1.009 & -0.74 & 3.67 & 0.97041 & 0.08090 & -0.01041\\
30 & 1.009 & -0.32 & -5.43 & 0.97029 & 0.08219 & -0.01027\\
35 & 1.002 & 2.21 & -19.12 & 0.96496 & 0.05313 & -0.00664\\
35 & 1.002 & 3.20 & -15.74 & 0.96542 & 0.05488 & -0.00686\\
35 & 1.002 & 4.56 & -14.01 & 0.96416 & 0.05063 & -0.00632\\
35 & 1.002 & 5.83 & -8.41 & 0.96528 & 0.05442 & -0.00680\\
\hline \hline
\end{tabular}
\label{table3}
\end{table*}
\subsection{Natural Potential}
As the last scalar field potential case, we analyze the
inflationary observables using the natural potential, which is
defined as
\begin{equation}\label{natpot}
V(\phi)=\Lambda ^4\left[1+\cos\left(\dfrac{\phi}{l}\right)\right],
\end{equation}
where $\Lambda$ and $l$ are constants with mass dimensions. For
successful inflation in such a potential, $l$ is scaled as the
Planck mass and $\Lambda$ is scaled as the grand unified mass
$M_{\rm GUT}\sim10^{16 }~{\rm GeV}$. With the natural potential,
inflation naturally takes place and a pseudo-Nambu-Goldstone boson
is a responsible field for it~\cite{natural, natcob, partmod}.
However with this potential, the predictions of GR inflation for
the tensor-to-scalar ratio is strongly disfavored by the joint
Planck, BK15 and BAO data~\eqref{planckdata2}, which motivates to
study inflation in the concept of $f(Q,T)$ formalism.

\begin{figure*}[t!]
\centering
\subfigure[]{
\includegraphics[width=0.31\textwidth]{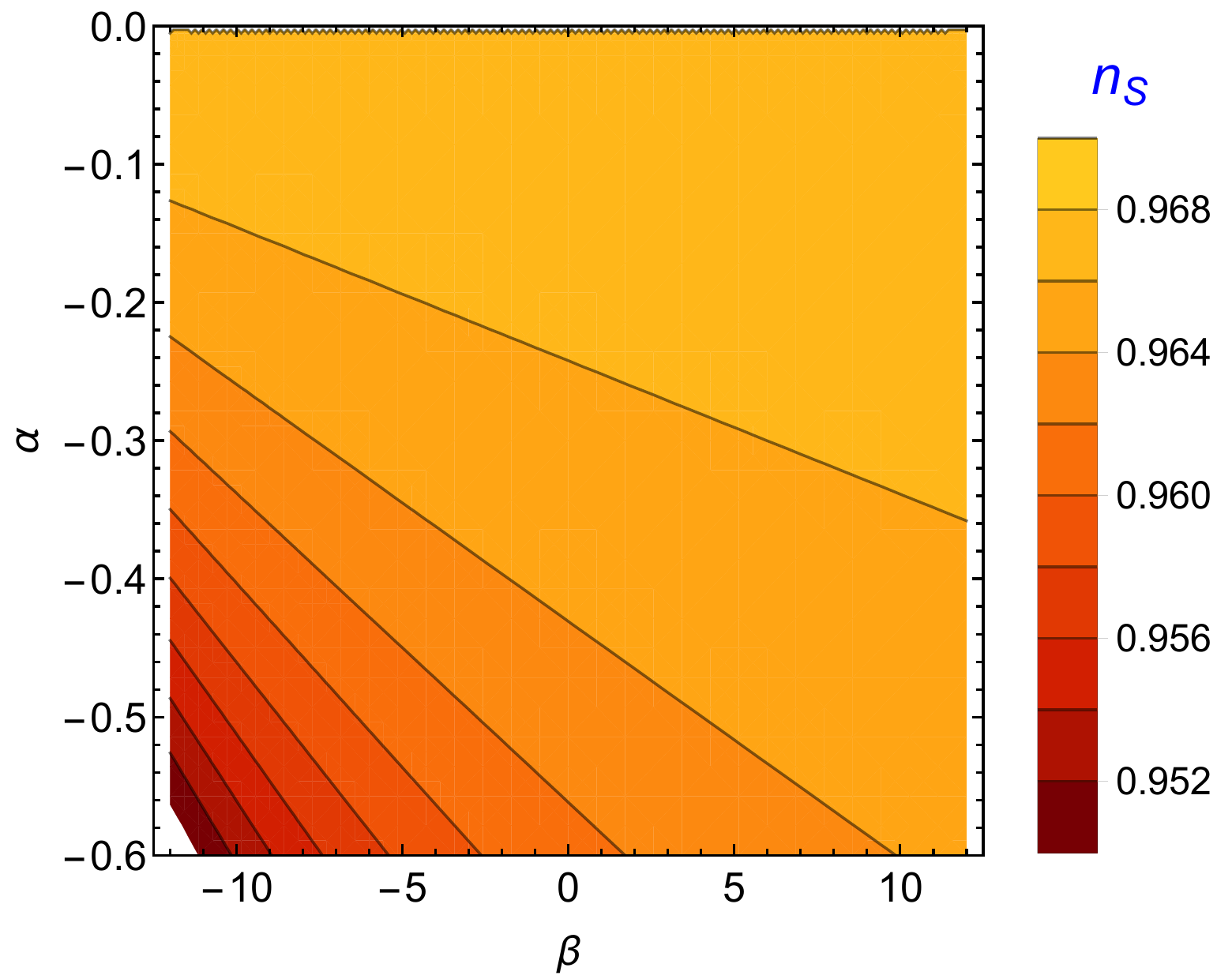}
\label{aa5}
}
\subfigure[]
{
\includegraphics[width=0.32\textwidth]{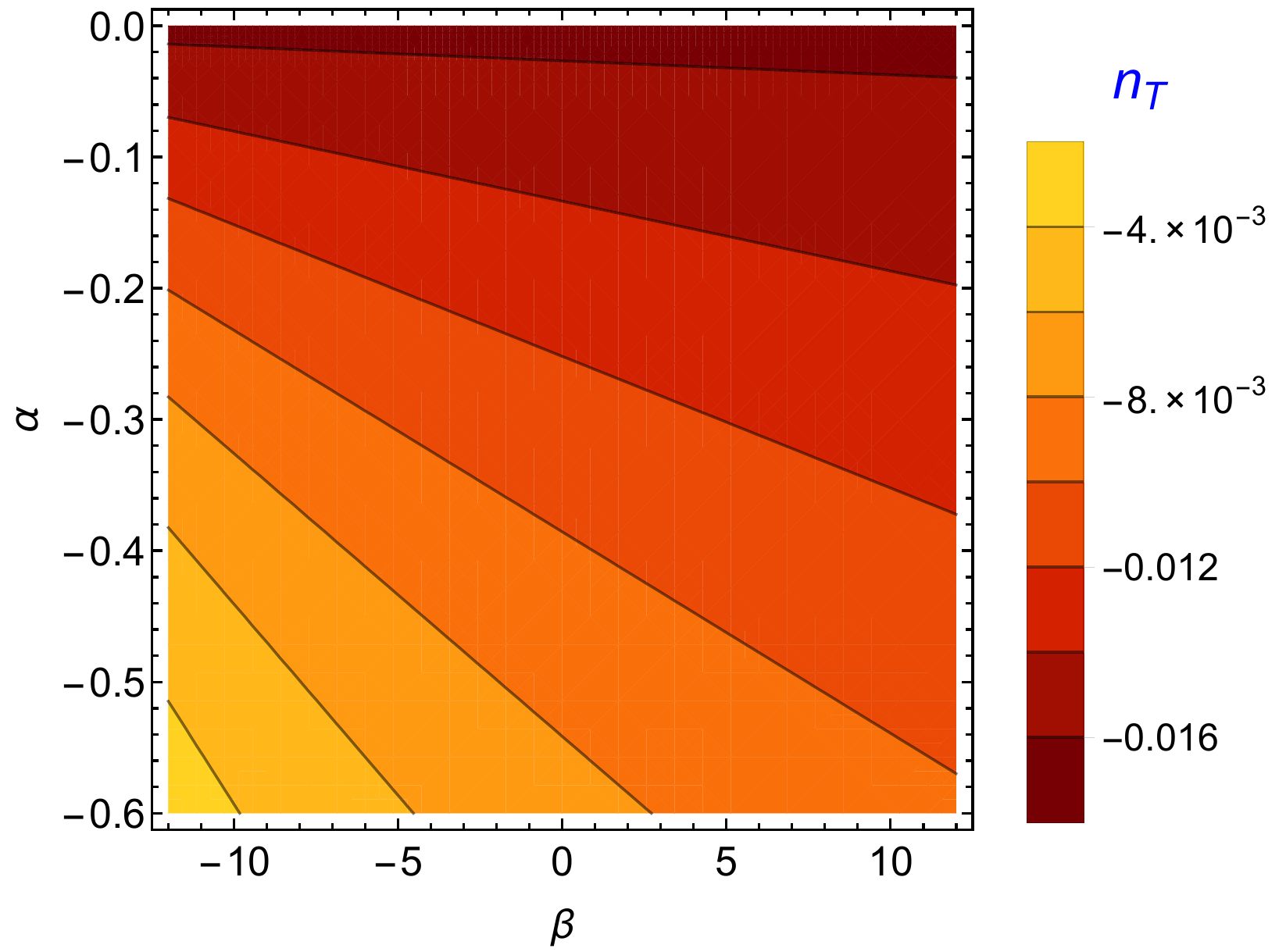}
\label{bb5}
}
\subfigure[]
{
\includegraphics[width=0.31\textwidth]{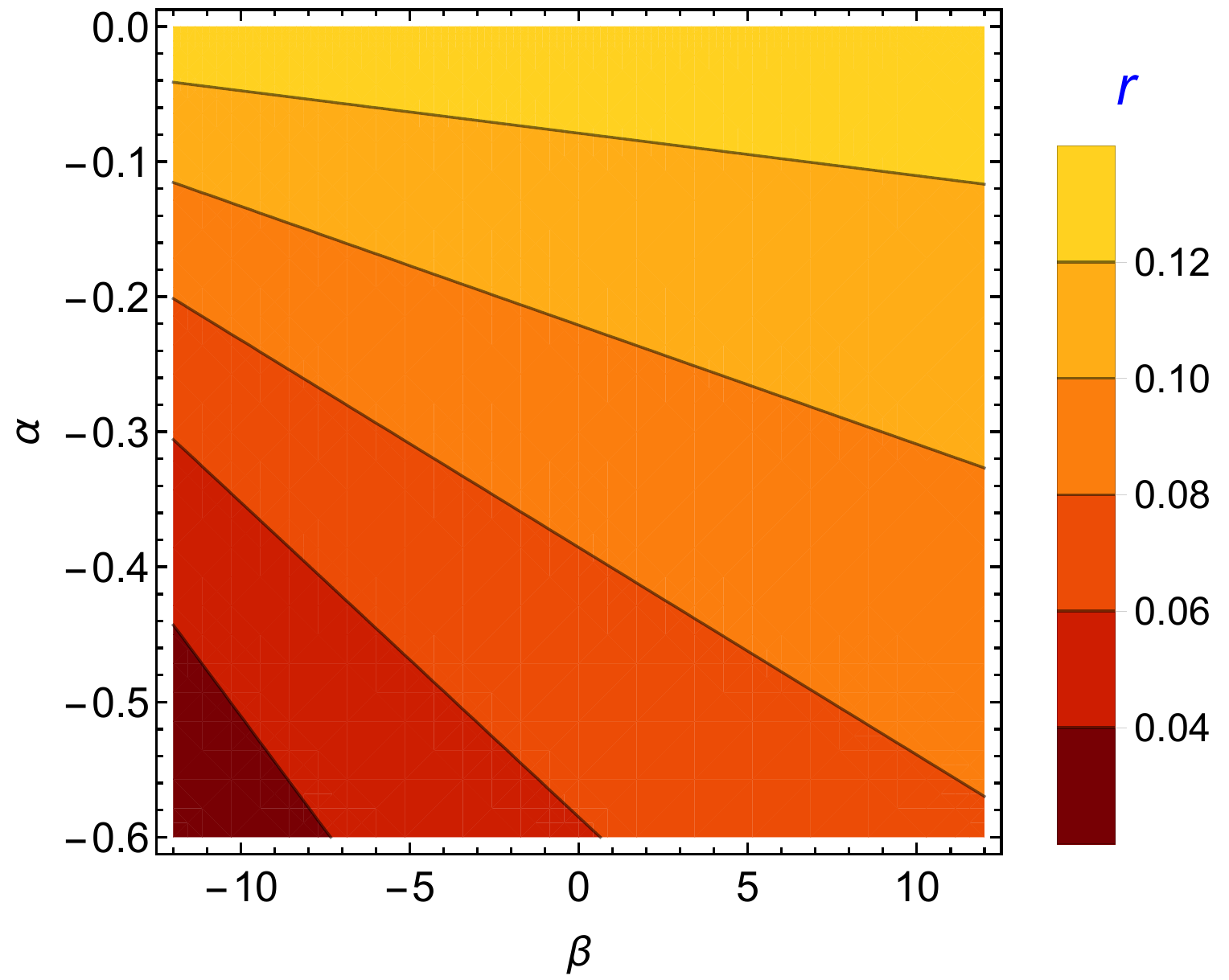}
\label{cc5}
}
\subfigure[]{
\includegraphics[width=0.31\textwidth]{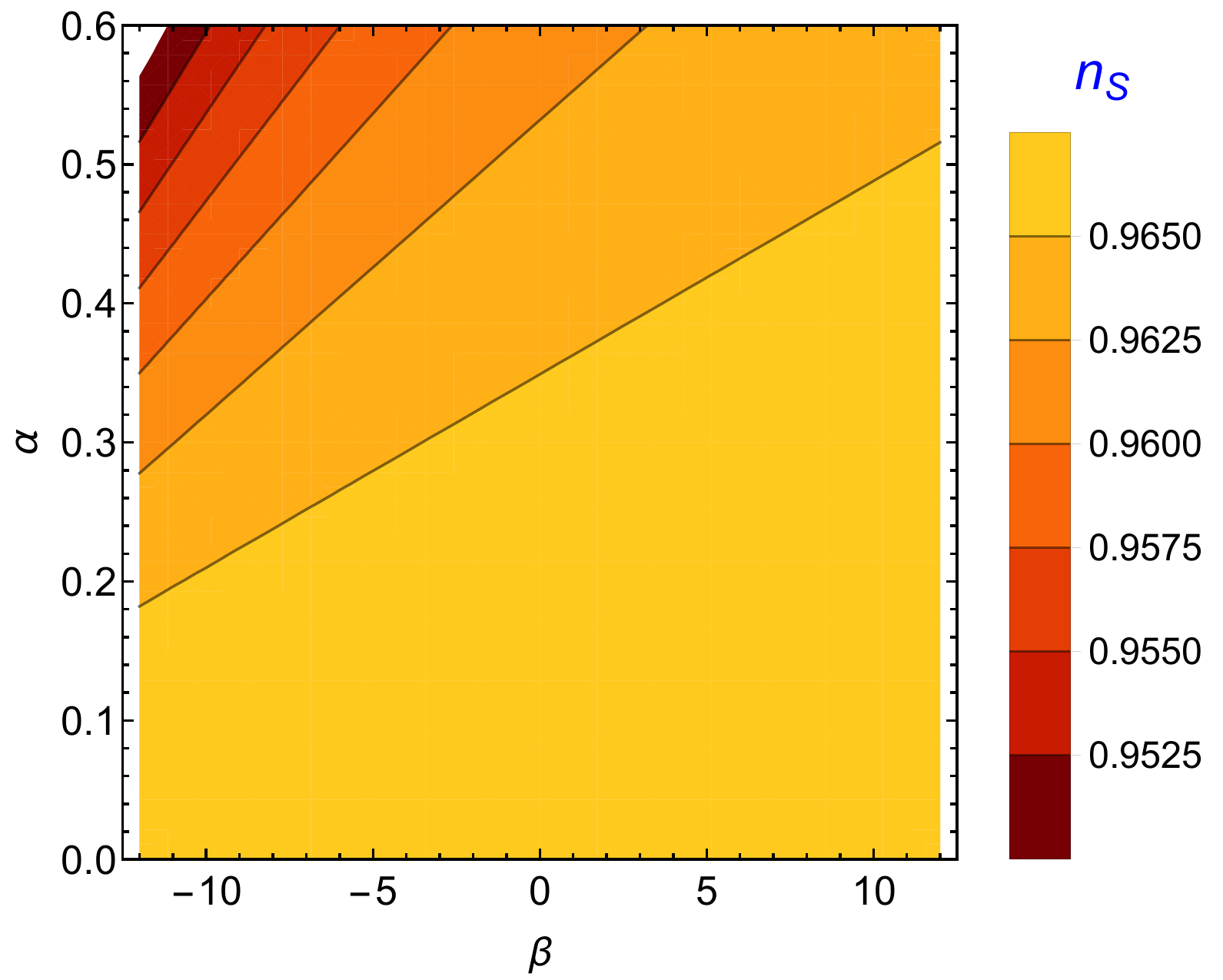}
\label{dd5}
}
\subfigure[]
{
\includegraphics[width=0.32\textwidth]{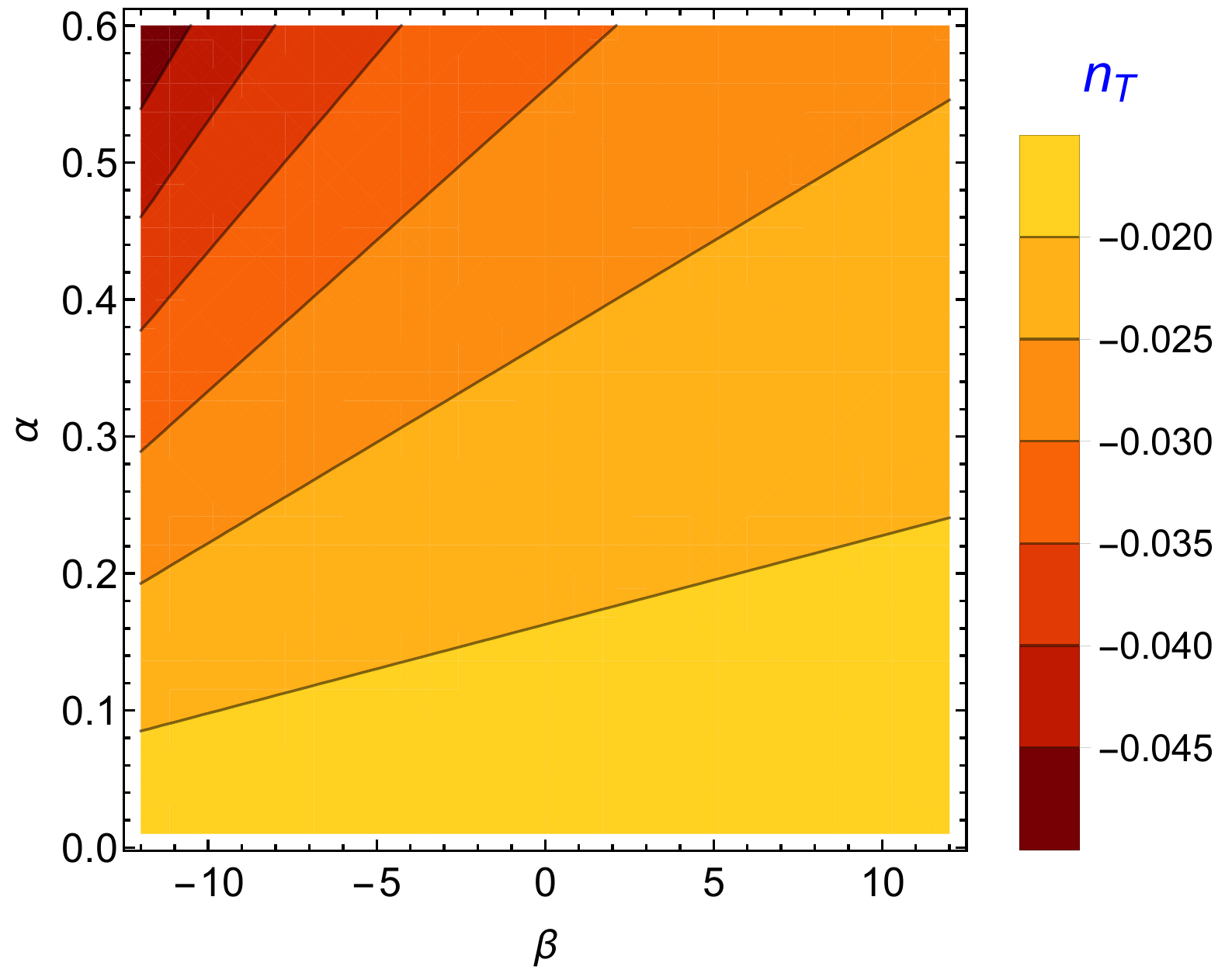}
\label{ee5}
}
\subfigure[]
{
\includegraphics[width=0.31\textwidth]{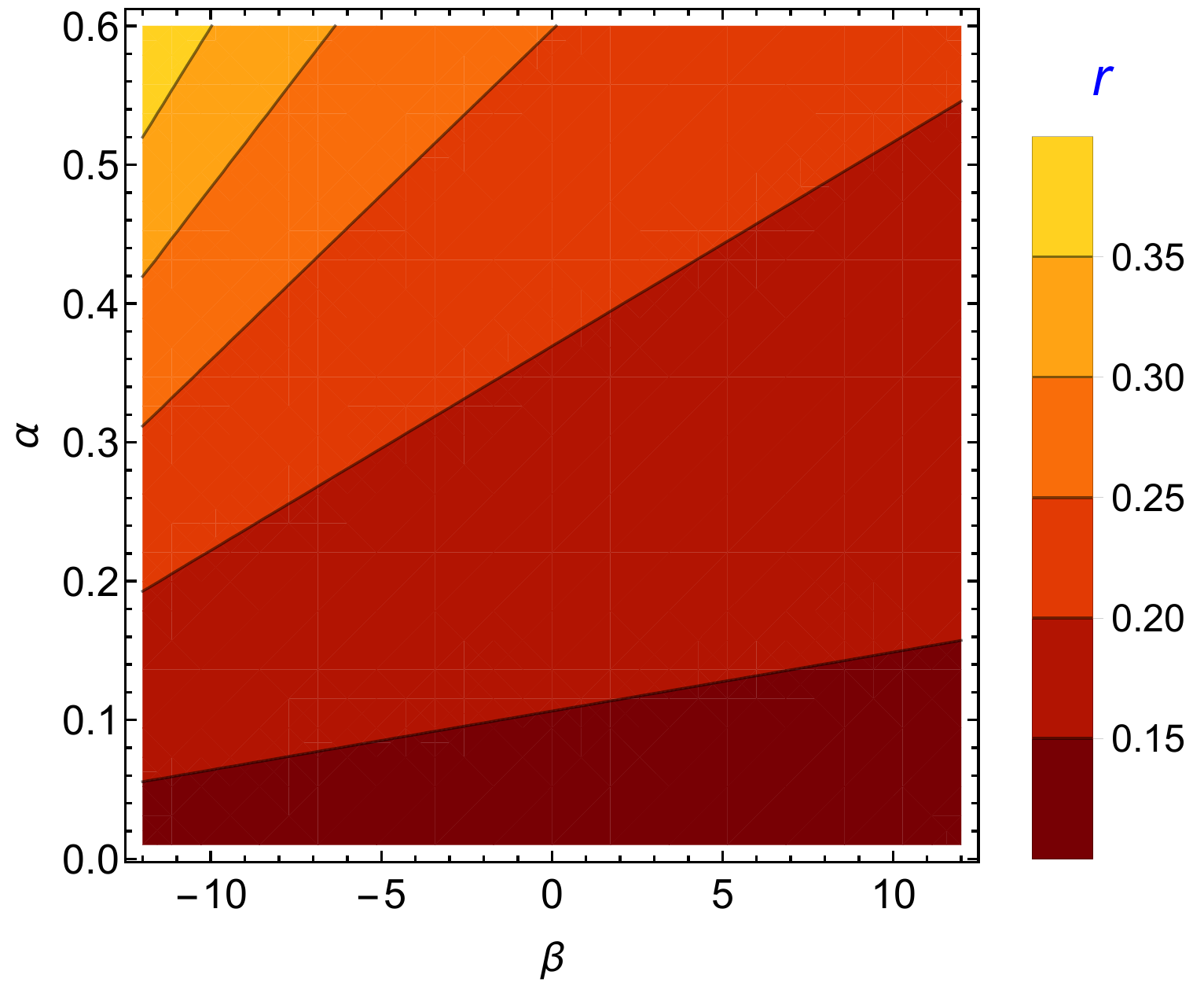}
\label{ff5}
}

\caption{[color online] Inflationary observables for the natural
potential as functions of parameters $\alpha$ and $\beta$. Top
figures are for $\alpha<0$, and bottom figures for $\alpha >0$.
All of these figures have been depicted for $l=5\,M_{\rm Pl}$ and
$N=60$.} \label{fig5}
\end{figure*}

\begin{table*}[ht]
\caption{Inflationary observables in the natural inflation for
some values of $l$ , $\alpha$ and $\beta$. Note that, for some of
the given values, the condition $-\alpha/(\kappa+\beta)>0$
is relaxed in order to consider further matter configurations.}
\centering
\begin{tabular}{c | c | c | c | c | c}
\hline \hline $l$ (in $M_{\rm Pl}$) & $\alpha$ & $\beta$ & $n_{\rm
S}$ & $r$ & $n_{\rm T}$ \\ [0.5ex] \hline
3 & -0.16 & -7.51 & 0.96069  & 0.05574 & -0.00697 \\
3 & -0.20 & -2.41 & 0.96106  & 0.05743 & -0.00717 \\
3 & -0.28 & 4.61 & 0.96025  & 0.05388 & -0.00673 \\
3 & -0.33 & 10.97 & 0.96062  & 0.05543 & -0.00692 \\
5 & -0.25 & -10.05 & 0.96418  & 0.07658 & -0.00957 \\
5 & -0.48 & 8.42 & 0.96488  & 0.08295 & -0.01036 \\
5 & -0.85 & 6.76 & 0.96000  & 0.05283 & -0.03283 \\
5 & -1.03 & 13.61 & 0.96003  & 0.05254 & -0.00662 \\
5 & 0.07 & -3.56 & 0.96683  & 0.14590 & -0.01823 \\
8 &  -0.86 & -5.43  & 0.96402   &  0.07530 &  -0.00941 \\
8 & -1.12  & 1.64  &  0.96426  & 0.07723  & -0.00965  \\
8 &  -1.32 & -5.64  &  0.96009  & 0.05322  & -0.00942  \\
8 &  -2.33 & 10.24  &  0.96045  & 0.05472  & -0.00684  \\
8 &  0.87 & -1.76  &  0.96481  & 0.19920  & -0.02490  \\
\hline
\hline
\end{tabular}
\label{table4}
\end{table*}
Similar to the previous cases, the slow-roll parameters for the
natural potential are
\begin{align}
\label{epsnat}&{\rm\epsilon}\approx\dfrac{-\alpha}{2\left(\kappa+\beta\right)l^2}\dfrac{\sin^2
\left(\dfrac{\phi}{l}\right)}{\left[1+\cos\left(\dfrac{\phi}{l}\right)\right]^2},\\
\label{etanat}&\eta
\approx\dfrac{\alpha}{\left(\kappa+\beta\right)l^2}\dfrac{\cos\left(\dfrac{\phi}{l}
\right)}{\left[1+\cos\left(\dfrac{\phi}{l}\right)\right]}.
\end{align}
Then, by using condition ${\rm \epsilon(\phi_{\rm end})}=1$, one
obtains $\phi_{\rm end}$ as
\begin{equation}\label{phiendnat}
\phi_{\rm end}\approx l\times\,\arccos\left[\dfrac{\alpha+2
l^2\left(\kappa+\beta\right)}{\alpha-2 l^2
\left(\kappa+\beta\right)}\right],
\end{equation}
where the argument of cosine must be\footnote{The first equality
is omitted because otherwise it makes $\epsilon =0$, and the
reason for omitting the second equality is given below
relation~\eqref{phinat}.}\
 $-1<
[\alpha+2 l^2\left(\kappa+\beta\right)]/[\alpha-2 l^2
\left(\kappa+\beta\right)]<1$ that leads once again to condition
$-\alpha/(\kappa+\beta)>0$. Moreover, the e-folding number for
this potential takes the form
\begin{equation}\label{natefold}
N\approx\dfrac{\left(\kappa+\beta\right)l^2}{\alpha}\ln\left[\dfrac{1-\cos\left(\dfrac{\phi}{l}
\right)}{1-\cos\left(\dfrac{\phi_{\rm end}}{l}\right)}\right].
\end{equation}
By substituting relation~\eqref{phiendnat} into
relation~\eqref{natefold}, one can express $\phi$ in terms of the
e-folding number as
\begin{equation}\label{phinat}
\phi\!\approx\!l\times\arccos\!\left\lbrace\!\dfrac{\alpha\!-2l^2\left(\kappa+\beta\right)\!\left[1\!-2\exp
\left(\dfrac{\alpha
N}{\left(\kappa+\beta\right)l^2}\right)\right]}{\alpha-2l^2\left(\kappa+\beta\right)}\!\right\rbrace\!,
\end{equation}
where $\cos\left(\phi_{\rm end}/l\right)\neq 1$ (that is why we
omitted the second equality in the argument of cosine below
relation~\eqref{phiendnat}). Under these considerations, the
slow-roll parameters are
\begin{equation}
\label{epspnat}{\rm
\epsilon}\approx\dfrac{\alpha\exp\left(\dfrac{\alpha
N}{\left(\kappa+\beta \right)l^2}\right)}{\alpha-2
l^2\left(\kappa+\beta\right)\left[1- \exp \left(\dfrac{\alpha
N}{\left(\kappa+\beta\right)l^2}\right)\right]},
\end{equation}
\begin{equation}
\label{etapnat}\eta\approx\dfrac{\alpha\left\lbrace \alpha-2l^2
\left(\kappa+\beta\right) \left[1-2\exp\left(\dfrac{\alpha
N}{\left(\kappa+\beta\right)l^2}\right)\right] \right\rbrace
}{2l^2\left(\kappa+\beta\right)\!\left\lbrace
\alpha\!-2l^2\left(\kappa+\beta
\right)\!\left[1\!-\!\exp\left(\dfrac{\alpha
N}{\left(\kappa+\beta\right)l^2}\right)\!\right]\right\rbrace}.
\end{equation}
Consequently, the inflationary observables obviously read
\begin{align}
\label{nspnat}&n_{\rm S}\approx
1-\dfrac{6\alpha\exp\left(\dfrac{\alpha
N}{\left(\kappa+\beta\right)l^2} \right)}{\alpha-2
l^2\left(\kappa+\beta\right)\left[1- \exp\left(\dfrac{\alpha N}
{\left(\kappa+\beta\right)l^2}\right)\right]}\nonumber\\
&+\dfrac{\alpha\left\lbrace \alpha-2l^2
\left(\kappa+\beta\right)\left[1-2\exp \left(\dfrac{\alpha
N}{\left(\kappa+\beta\right)l^2}\right)\right] \right\rbrace }{l^2
\left(\kappa+\beta\right)\!\left\lbrace
\alpha\!-2l^2\left(\kappa+\beta\right)\!\left[1\!-\exp
\left(\dfrac{\alpha
N}{\left(\kappa+\beta\right)l^2}\right)\right]\right\rbrace},
\end{align}
\begin{align}
\label{ntpnat}&n_{\rm
T}\approx\dfrac{-2\alpha\exp\left(\dfrac{\alpha
N}{\left(\kappa+\beta \right)l^2}\right)}{\alpha-2
l^2\left(\kappa+\beta\right)\left[1- \exp\left(\dfrac{\alpha N}
{\left(\kappa+\beta\right)l^2}\right)\right]},
\end{align}
\begin{align}
\label{rnat}&r\approx\dfrac{16\alpha\exp\left(\dfrac{\alpha
N}{\left(\kappa+\beta\right)l^2} \right)}{\alpha-2
l^2\left(\kappa+\beta\right)\left[1- \exp\left(\dfrac{\alpha
N}{\left(\kappa+\beta\right)l^2}\right)\right]}.
\end{align}
As is clear, with the natural potential, these relations depend on
the e-folding number and on a single combination of the other
parameters, i.e. $-l^2 \(\kappa+\beta\)/\alpha$, with respect to
GR. The scaling change of the mass scale $l$, i.e.
\begin{equation}\label{lScaling}
l^2\kappa \rightarrow \dfrac{l^2 (\kappa+\beta)}{-\alpha},
\end{equation}
also stems from relation~\eqref{modified kapp} that leads to the
relations resulting from GR with a minimally coupled scalar field
(see the Appendix). Such a modification can cause some differences
compared to the results of the GR case. For instance, the natural
inflation can provide~\cite{Planck:2018jri} consistent results
with the Planck 2018 data in the case of GR if $0.3 <
\log_{10}(l/{\rm M_{Pl}}) < 2.5$ at $95\%\ {\rm CL}$. In this case
of linear $f(Q, T)$ gravity, relation \eqref{lScaling} can impose
different restrictions on the corresponding $l$ value, i.e. $0.3+F
< \log_{10}(\tilde{l}/{\rm M_{Pl}}) < 2.5+F$ where
$F\equiv\log_{10}\sqrt{(\kappa+\beta)/(-\alpha\kappa)}$, where we
have set $l^2\kappa \rightarrow \tilde{l}^2\kappa$.

Fig.~\ref{fig5} indicates the scalar spectral index, the tensor
spectral index, and the tensor-to-scalar ratio for $N=60$,
$l=5\,M_{\rm Pl}$ with different values of $\beta$ and $\alpha$
for the natural potential. Also, Table~\ref{table4} shows the
inflationary observables for different values of $l$, $\alpha$ and
$\beta$. The results indicate that, for some ranges of free
parameters within the context of the linear functional form of the
$f(Q, T)$ gravity with the natural potential, the inflationary
observables are in very good agreement with the observational data
obtained from the Planck satellite, i.e., data~\eqref{planckdata}.
Furthermore, with the natural potential, it is interesting that,
for some negative values of $\alpha$ and appropriate $\beta$
(consistent with the related condition), our findings make $r$
even more restrictive and (within the given limits of $n_S$) in
good agreement with the joint Planck, BK15 and BAO data, i.e.
relation \eqref{planckdata2}, in contrast to the results obtained
from GR in this case.

\section{Conclusions}
In recent decades, a wide range of investigations has been
performed to describe the dynamics of the Universe. In this
regard, the results of studies somehow reflect the fact that the
standard model of cosmology derived from GR offers the best
description. Nevertheless, the CMB observables contain very
important information about the formation and evolution of the
Universe, and some fundamental concepts, such as the flatness and
horizon problems, still have remained as open issues within the
context of the standard model of cosmology. To address these
shortcomings, further research on cosmological inflation appears
to be needed in the earliest stages of the Universe. However,
although GR has made accurate predictions to describe the
cosmological phenomena, it is undesirable to justify the effect of
dark sectors on the dynamics of the Universe in a way that is well
consistent with the observational data. For this purpose, studying
alternative models of gravity can be a good motivated.

In this work, we have investigated the cosmological inflation
within the context of $f(Q, T)$ gravity. To perform this task,
first we have described the simplest theoretical framework for
cosmological inflation based on GR with an isotropic and
homogeneous scalar field known as inflaton. Then, by assuming the
spatially flat FLRW spacetime and a linear barotropic equation of
state, we have introduced the slow-roll parameters in terms of the
Hubble parameter and its time derivatives. For inflation to occur
and have a sufficient time scale as well as timely end of
inflation, the conditions $|{\rm\epsilon_n}|\ll 1$ and
${\rm\epsilon}=1$ must be satisfied. In addition, to calculate the
inflationary parameters in the presence of the scalar field, we
have derived the potential representation of slow-roll parameters.
We have also applied the slow-roll conditions in the related
calculations.

By considering the energy-momentum tensor as a perfect fluid, we
have calculated the modified Friedman equations, the effective
pressure and energy densities, and the evolution of energy density
extracted from $f(Q, T)$ gravity. Furthermore, by choosing a
linear combination of $Q$ and $T$, we have shown that the
evolution of phantom and quintessence dominated era, that unified
two-phase of acceleration of the Universe in the early and
late-time, can be achieved in $f(Q,T)$ gravity theory. Also, we
have modeled the inflationary scenario for the linear $f(Q, T)$
gravity, and have calculated the slow-roll parameters, the scalar
spectral index, the tensor spectral index, and the
tensor-to-scalar ratio. The results show that the inflationary
quantities within the context of the linear $f(Q, T)$ gravity are
independent of the e-folding number and the free parameter
$\alpha$. Instead, they depend only on the parameters $\beta$ and
$w$. In addition, we have indicated that with the values ($w=-1$
with $\beta\neq -\kappa/2$) and/or ($\beta=-\kappa$ with $w\neq
1$), no inflation could have occurred in the early Universe, and
in fact these values represent the expansion phase of de~Sitter.
Also, for the values ($w=-1$ with $\beta = -\kappa/2$) and/or
($\beta=-\kappa$ with $w= 1$), the solution to the field equations
is the Minkowski metric with no inflation.

Furthermore, we have investigated the slow-roll inflation in the
presence of a scalar field within the context of the linear $f(Q,
T)$ gravity, and have calculated the corresponding inflationary
observables. We have also computed the inflationary observables
for this model with three different cases of inflationary
potentials, namely the power-law, the hyperbolic and the natural
potentials.

For the power-law potential, the theoretical results indicate that
the inflationary observables depend only on the power of the
scalar field, $n$, and the e-folding number. We have specified
that this potential for the linear $f(Q, T)$ gravity does~not lead
to further correction to the inflationary observables compared to
those ones extracted from GR.

For the case of hyperbolic potential, by fixing the e-folding
number and the two parameters of the potential, the related
inflationary observables obviously depend only on the free
parameters of the linear form, i.e. $\alpha$ and $\beta$. The
theoretical results demonstrate that, by considering the
hyperbolic potential for the linear $f(Q, T)$ gravity, the
appropriate values of inflationary observables can be found to be
in good agreement with the observational data obtained from the
Planck satellite for some ranges of negative and positive values
of $\alpha$. However, while relaxing the condition
$-\alpha/\(\kappa+\beta\)>0$, the positive values of $\alpha$ give
even consistent results with the joint Planck, BK15 and BAO data.
Furthermore, the contribution of linear $f(Q,T)$ gravity can
impose some differences in the restrictions on the parameter of
the hyperbolic potential compared to the results obtained for the
GR case.

Finally, for the natural potential, again by fixing the e-folding
number and the parameter of the potential, the related
inflationary observables obviously depend only on the free
parameters of the linear form. The theoretical results indicate
that, within the context of the linear $f(Q, T)$ gravity with the
natural potential, the obtained inflationary parameters are
not~only in good agreement with the Planck data but (in contrast
to the results obtained from GR in this case) also well consistent
with the joint Planck, BK15 and BAO observational data, which
impose tighter constraint on the value of the tensor-to-scalar
ratio. These results justify the use of the $f(Q,T)$ gravity.
Furthermore, the contribution of linear model of $f(Q,T)$ gravity
can also impose some differences in the restrictions on the
parameter of the natural potential compared to the results
obtained for the GR case.

\section*{Appendix}
The symmetric teleparallel gravity has some generalizations, one
of which is known as $f(Q,T)$ gravity. In this appendix, we review
some prerequisites related to this theory of gravity.

In differential geometry, any general connection (e.g.
$\Gamma^{\alpha}{}_{\mu\nu}$) can obviously be decomposed into
three independent components as
\begin{equation}\label{affine}
\Gamma^{\alpha}{}_{\mu\nu}=\{^{\alpha}{}_{\mu\nu}\}+K^{\alpha}{}_{\mu\nu}+L^{\alpha}{}_{\mu\nu},
\end{equation}
where $\{^{\alpha}{}_{\mu\nu}\}$, $K^{\alpha}{}_{\mu\nu}$ and
$L^{\alpha}{}_{\mu\nu}$ respectively are the Christoffel symbol,
the contorsion tensor and disformation tensor defined
as\footnote{We follow the sign convention of Ref.~\cite{MTW} for
the covariant derivative, e.g. $\nabla_{\boldsymbol \gamma}\,
g_{\mu\nu}=\partial_\gamma
g_{\mu\nu}-\Gamma^\alpha{}_{\mu{\boldsymbol \gamma}}\,
g_{\alpha\nu}-\Gamma^\alpha{}_{\nu{\boldsymbol \gamma}}\,
g_{\mu\alpha}$, i.e. the differentiation index $\gamma$ comes
second in the lower indices of the connection.}\
\begin{eqnarray}
\label{leci}\{^{\alpha}{}_{\mu\nu}\}&=&\dfrac{1}{2}g^{\alpha\beta}(\partial_{\mu}
g_{\beta\nu}
 +\partial _{\nu}g_{\beta\mu}-\partial_{\beta} g_{\mu\nu}),\cr
 K^{\alpha}{}_{\mu\nu}&=&\dfrac{1}{2}\tau^{\alpha}{}_{\mu\nu}+\tau_{(\mu\nu)}{}^{\alpha},\cr
 L^{\alpha}{}_{\mu\nu}&\equiv &-\dfrac{1}{2}g^{\alpha\beta}\left(Q_{\mu\beta\nu}
 +Q_{\nu\beta\mu}-Q_{\beta\mu\nu}\right),
\end{eqnarray}
where $\tau^{\alpha}{}_{\mu\nu}= 2\Gamma^{\alpha}{}_{[\mu\nu]}$ is
the torsion tensor and $Q_{\alpha\mu\nu}\equiv\nabla_{\alpha}\,
g_{\mu\nu}$ is the nonmetricity tensor. In addition, the
nonmetricity scalar is defined as
\begin{equation}\label{invariant}
Q\equiv -g^{\mu
\nu}\left(L^{\alpha}{}_{\beta\mu}L^{\beta}{}_{\nu\alpha}-L^{\alpha}{}_{\beta\alpha}L^{\beta}{}_{\mu\nu}\right).
\end{equation}

It is well-known that for a zero Riemann tensor with torsionless
case, i.e. a flat manifold, there always exists an adapted
coordinate system, in which the connection is zero
everywhere\rlap.\footnote{It has been shown that the spatially
flat FLRW spacetime admits three distinct connections and only one
of those can become zero at the Cartesian coordinate system with
which the line-element is used. The other two lead to a $Q$ that
depends also on the connection~\cite{DAmbrosio:2021pnd,Hohmann}.}\
 Hence, in such a system, relation~\eqref{affine} leads to
\begin{equation}
L^{\alpha}{}_{\mu\nu}\mathrel{\mathop=^{*}}-\{^{\alpha}{}_{\mu\nu}\}.
\end{equation}
Subsequently, the nonmetricity scalar~\eqref{invariant} in such a
system is
\begin{equation}
Q\mathrel{\mathop=^{*}}-g^{\mu
\nu}\Bigl(\{^{\alpha}{}_{\beta\mu}\}\{^{\beta}{}_{\nu\alpha}\}-\{^{\alpha}{}_{\beta\alpha}\}\{^{\beta}{}_{\mu\nu}\}\Bigr)
\end{equation}
that is exactly equal to minus the effective Einstein-Hilbert
Lagrangian (which is~not a scalar in this form). Hence, the linear
version of $f(Q)$ gravity in the absence of the boundary terms is
dynamically equivalent to results of the Einstein-Hilbert action
and provides another geometrical formalism for GR. Accordingly,
the linear version $f(Q,T)=\alpha\, Q+\beta\, T$ is dynamically
equivalent to the linear version $f(R,T)=\alpha\, R+\beta\, T$
theory that is claimed to be the GR case with a slight
modification in the matter
content~\cite{Fisher2019,HMoraes,Fisher2020}.

Indeed, the equations of motion of the linear $f(Q,T)=\alpha\,
Q+\beta\, T$, wherein $T_{\mu\nu}$ is the usual energy-momentum
tensor of a scalar field (e.g. $\phi$), become those of GR with
another minimally coupled scalar field (e.g. $\psi$), where
\begin{equation}\label{minimally field}
\psi=\sqrt{\frac{\kappa+\beta}{-\alpha\kappa}}\,\phi ,
\end{equation}
with a potential
\begin{equation}\label{minimally potential}
\tilde{V}(\psi)=\frac{\kappa+2\beta}{-\alpha\kappa}V(\phi).
\end{equation}
However, this equivalence is a matter of debate, and we intend to
justify the use of the linear $f(Q,T)$ gravity by providing a
comparison showing possible differences with that obtained from
the GR case.

\section*{ACKNOWLEDGMENTS}
The authors thank the Research Council of Shahid Beheshti
University.


\end{document}